\documentclass[pdflatex,sn-nature]{sn-jnl}

\usepackage{times}
\usepackage{graphicx}%
\usepackage{multirow}%
\usepackage{amsmath,amssymb,amsfonts,bm}%
\usepackage{amsthm}%
\usepackage{mathrsfs}%
\usepackage[title]{appendix}%
\usepackage{xcolor}%
\usepackage{textcomp}%
\usepackage{manyfoot}%
\usepackage{booktabs}%
\usepackage{algorithm}%
\usepackage{algorithmicx}%
\usepackage{algpseudocode}%
\usepackage{listings}%
\usepackage{upgreek}

\def\bra#1{\mathinner{\langle{#1}|}} 
\def\ket#1{\mathinner{|{#1}\rangle}}

\begin{document}

\title[Article Title]{Multimode Ultrastrong Coupling in Three-Dimensional Photonic-Crystal Cavities}
%

\author[1,2]{\fnm{Fuyang} \sur{Tay}}
\author[1]{\fnm{Ali} \sur{Mojibpour}}
\author[1]{\fnm{Stephen} \sur{Sanders}}
\author[3,4]{\fnm{Shuang} \sur{Liang}}
\author[5]{\fnm{Hongjing} \sur{Xu}}
\author[6]{\fnm{Geoff} \sur{C. Gardner}}
\author[1,7]{\fnm{Andrey} \sur{Baydin}}
\author[3,4,6,8]{\fnm{Michael} \sur{J. Manfra}}
\author[1]{\fnm{Alessandro} \sur{Alabastri}}
\author[9]{\fnm{David} \sur{Hagenm\"uller}}
\author*[1,5,7,10]{\fnm{Junichiro} \sur{Kono}}\email{kono@rice.edu}

\affil[1]{\orgdiv{Department of Electrical and Computer Engineering}, \orgname{Rice University}, \orgaddress{\city{Houston}, \state{Texas} \postcode{77005}, \country{USA}}}
\affil[2]{\orgdiv{Applied Physics Graduate Program, Smalley--Curl Institute}, \orgname{Rice University}, \orgaddress{\city{Houston}, \state{Texas} \postcode{77005}, \country{USA}}}
\affil[3]{\orgdiv{Department of Physics and Astronomy}, \orgname{Purdue University}, \orgaddress{\city{West Lafayette}, \state{Indiana} \postcode{47907}, \country{USA}}}
\affil[4]{\orgdiv{Birck Nanotechnology Center}, \orgname{Purdue University}, \orgaddress{\city{West Lafayette}, \state{Indiana} \postcode{47907}, \country{USA}}}
\affil[5]{\orgdiv{Department of Physics and Astronomy}, \orgname{Rice University}, \orgaddress{\city{Houston}, \state{Texas} \postcode{77005}, \country{USA}}}
\affil[6]{\orgdiv{School of Electrical and Computer Engineering}, \orgname{Purdue University}, \orgaddress{\city{West Lafayette}, \state{Indiana} \postcode{47907}, \country{USA}}}
\affil[7]{\orgdiv{Smalley--Curl Institute}, \orgname{Rice University}, \orgaddress{\city{Houston}, \state{Texas} \postcode{77005}, \country{USA}}}
\affil[8]{\orgdiv{School of Materials Engineering}, \orgname{Purdue University}, \orgaddress{\city{West Lafayette}, \state{Indiana} \postcode{47907}, \country{USA}}}
\affil[9]{\orgdiv{CESQ-ISIS (UMR 7006)}, \orgname{Universit\'e de Strasbourg and CNRS}, \orgaddress{\city{Strasbourg}, \postcode{67000}, \country{France}}}
\affil[10]{\orgdiv{Department of Materials Science and NanoEngineering}, \orgname{Rice University}, \orgaddress{\city{Houston}, \state{Texas} \postcode{77005}, \country{USA}}}

\affil{}{}

\abstract{Recent theoretical studies have highlighted the role of spatially varying cavity electromagnetic fields in exploring novel cavity quantum electrodynamics (cQED) phenomena, such as the potential realization of the elusive Dicke superradiant phase transition. One-dimensional photonic-crystal cavities (PCCs), widely used for studying solid-state cQED systems, have uniform spatial profiles in the lateral plane. Three-dimensional (3D) PCCs, which exhibit discrete in-plane translational symmetry, overcome this limitation, but fabrication challenges have hindered the achievement of strong coupling in 3D-PCCs. Here, we report the realization of multimode ultrastrong coupling in a 3D-PCC at terahertz frequencies. The multimode coupling between the 3D-PCC's cavity modes and the cyclotron resonance of a Landau-quantized two-dimensional electron gas in GaAs is significantly influenced by the spatial profiles of the cavity modes, leading to distinct coupling scenarios depending on the probe polarization. Our experimental results are in excellent agreement with a multimode extended Hopfield model that accounts for the spatial inhomogeneity of the cavity field. Guided by the model, we discuss the possible strong ground-state correlations between different cavity modes and introduce relevant figures of merit for the multimode ultrastrong coupling regime. Our findings emphasize the importance of spatially nonuniform cavity mode profiles in probing nonintuitive quantum phenomena expected for the ground states of cQED systems in the ultrastrong coupling regime.}


\maketitle

\section*{Introduction}
\label{sec:intro}

The interaction between a two-level system and a spatially uniform cavity electric field represents the most fundamental model in the field of cavity quantum electrodynamics (cQED). This model has been extensively studied over the past century and has provided profound insights into the nature of light--matter interactions in confined geometries. Recently, significant phenomena have been predicted to emerge when the cavity electromagnetic field is spatially nonuniform~\cite{NatafEtAl2019PRL,AndolinaEtAl2020PRB,GuerciEtAl2020PRL}. For example, the Dicke superradiant phase transition (SRPT)~\cite{HeppLieb1973AP,WangHioe1973PRA}, which is expected to occur when the light--matter coupling strength is comparable to the photon frequency but prohibited by a no-go theorem in the case of a uniform field~\cite{NatafCiuti2010NC,AndolinaEtAl2019PRB}, may occur when a spatially varying electromagnetic field is involved. The Dicke SRPT becomes possible when the cavity field strongly couples with a paramagnetic instability at finite wave vectors, which is only possible with a spatially varying electric field~\cite{NatafEtAl2019PRL,AndolinaEtAl2020PRB,GuerciEtAl2020PRL,ManzanaresEtAl2022PRB}.

Fabry--P\'erot cavities with Bragg mirrors, or one-dimensional photonic crystal cavities (1D-PCCs), are commonly employed to investigate solid-state cQED with quantum wells~\cite{DengEtAl2002S,GibbsEtAl2011NP,ZhangEtAl2016NP,LiEtAl2018NP}. These purely dielectric cavities offer a significant advantage in terms of high quality factors~\cite{GibbsEtAl2011NP,ZhangEtAl2016NP,LiEtAl2018NP}. However, the cavity mode profile in 1D-PCCs is spatially uniform in the plane perpendicular to the stacking direction. Three-dimensional (3D) PCCs overcome this limitation by exhibiting discrete in-plane translational invariance~\cite{NodaEtAl2000S,NodaEtAl2007NP,IshizakiEtAl2013NP}. This allows cavity photons to couple with matter excitations with a set of in-plane reciprocal lattice vectors. Coherent radiative hopping and significant changes in exciton dispersion have been predicted to occur in 3D-PCCs operating in the strong coupling regime~\cite{YangJohn2007PRB}. Nevertheless, fabrication challenges have prevented the realization of extreme regimes of light--matter interactions in 3D-PCCs, including the strong and ultrastrong coupling~\cite{Forn-DiazEtAl2019RMP,FriskKockumEtAl2019NRP} (USC) regimes that necessitate high quality factors and strong electric field confinement~\cite{VosWoldering2015LLaLRaPPS}.

In the present work, we demonstrate multimode USC in a terahertz (THz) 3D-PCC. The cavity modes of the 3D-PCC are simultaneously coupled to the cyclotron resonance (CR) of an ultrahigh-mobility two-dimensional electron gas (2DEG) in GaAs. Our 3D-PCC, which is nearly inversion-symmetric in the stacking direction, exhibits two degenerate cavity modes with different spatial profiles in the 2DEG plane depending on their polarization. Although the dipole approximation remains valid (the spatial profiles of the cavity modes are nearly uniform on the scale of the electron cyclotron orbits), the spatial variation of the cavity field within a unit cell of the 3D-PCC significantly affects the hybridization between the cavity modes through the CR, leading to the emergence of novel physics. Two distinct multimode scenarios in which different cavity modes are either coupled or decoupled via the CR are observed in a single device, by rotating the polarization of the probe. Our experimental results show excellent agreement with numerical simulations and calculations based on an extended Hopfield model, which takes into account the spatial variation of the cavity field. We discuss possible correlations between the cavity modes in the ground state of the system as indicated by the model and present relevant figures of merit for multimode systems with USC. This work represents the first study to explore light--matter interactions beyond the weak coupling regime in 3D-PCCs, and highlights the importance of spatial variation of cavity mode profiles in cQED.


\section*{Results}
\subsection*{Polarization-dependent multimode coupling in a 3D-PCC}\label{sec:3dpcc}
\begin{figure}[ht]%
\centering
\includegraphics[width=0.9\textwidth]{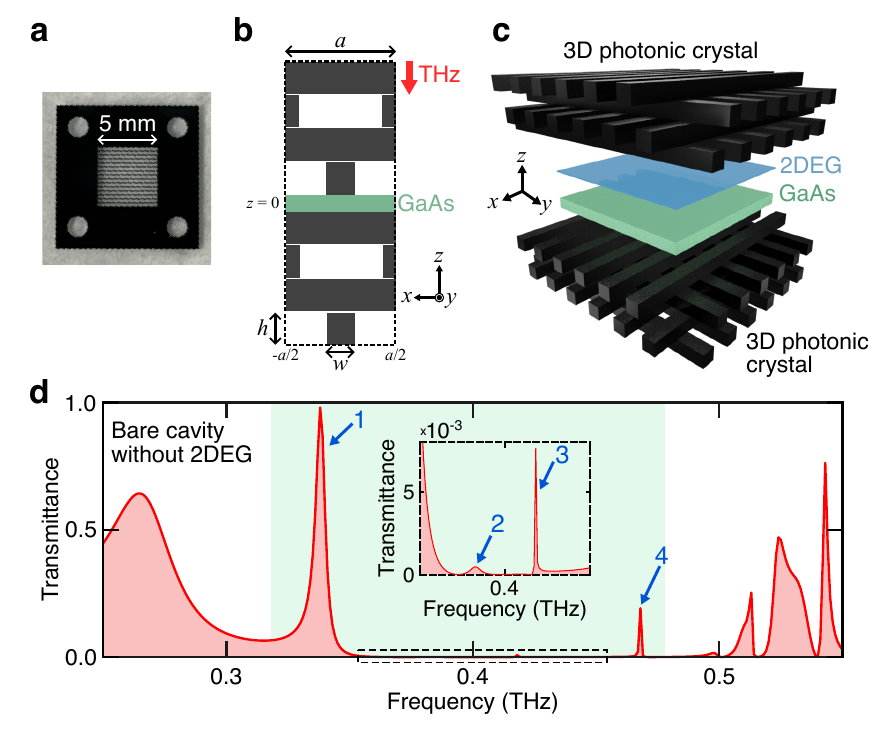}
\caption{\textbf{The 3D-PCC.} \textbf{a},~A photo of a fabricated architecture containing the designed woodpile structure. \textbf{b},~Side view of the unit cell of the 3D-PCC. A GaAs wafer is sandwiched by a pair of 3D photonic crystals, which consists of four layers with a rod array in the $z$ direction. Each silicon rod had a width of $w=0.26a$ and a height of $h=0.3a$ where $a=333$\,$\upmu$m is the lattice constant. The rods in each successive layer of the 3D photonic crystal are rotated by 90$^\circ$. The rods in the third and fourth layers are displaced by $0.5a$ relative to the rods in the first and second layers, respectively. The THz radiation propagates along the $-z$ direction. \textbf{c},~Schematic of a 2DEG embedded in a 3D woodpile cavity. \textbf{d},~Transmittance spectrum of the bare 3D-PCC from numerical simulations. The shaded region corresponds to the photonic band gap computed from the band structure of an infinite woodpile lattice (in all directions) without defect. The blue arrows mark the doubly degenerate cavity modes (with respect to the two polarizations) that are labeled by $p=1,2,3,4$ throughout the paper. The inset is a magnified view of the spectrum.}
\label{fig:schematic}
\end{figure}

3D photonic crystals are dielectric structures that exhibit periodic refractive index modulations in all three dimensions. Here, we consider the woodpile structure where the rod arrays are stacked in an alternating orthogonal pattern~\cite{HoEtAl1994SSC,SozuerDowling1994JMO}. We used photolithography and deep reactive ion etching to fabricate a rod array in silicon wafers (Fig.~\ref{fig:schematic}a, see Methods for more details). A woodpile lattice~\cite{IshizakiNoda2009N,IshizakiEtAl2013NP} is formed by stacking patterned silicon wafers sequentially (Fig.~\ref{fig:schematic}b). To induce photonic modes in the photonic band gap, a planar defect (a 60-$\upmu$m-thick GaAs layer) was inserted at the center of the woodpile lattice, which breaks discrete translational invariance in the stacking direction $z$ (Fig.~\ref{fig:schematic}b). We performed numerical simulations with COMSOL Multiphysics to calculate the transmittance spectrum of the woodpile structure under normal incidence, as shown in Fig.~\ref{fig:schematic}d. To maintain high peak amplitude, only four silicon layers were placed on each side of the GaAs defect layer. Four cavity modes can be observed within the photonic band gap. Those cavity modes exhibit finite quality factors that depend on their position in the band gap: The closer they are to the center of the band gap, the larger their quality factor is. Specifically, the quality factors of the modes with frequencies $\omega_{p=1}/2\pi=338$\,GHz, $\omega_{p=2}/2\pi=382$\,GHz, $\omega_{p=3}/2\pi=417$\,GHz, and $\omega_{p=4}/2\pi=468$\,GHz, as computed by finite difference time domain (FDTD) simulations, are 72, 70, 6200, and 1540, respectively.

\begin{figure}[ht!]%
\centering
\includegraphics[width=0.85\textwidth]{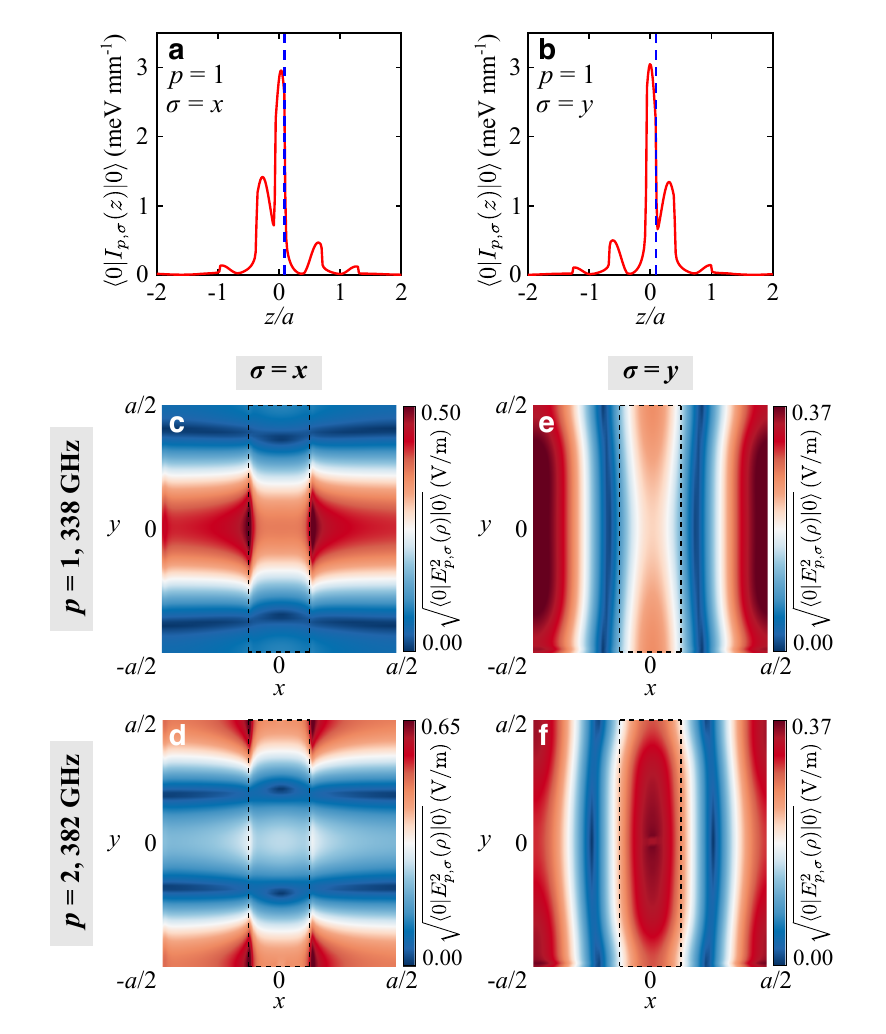}
\caption{\textbf{Polarization-dependent mode profiles.} \textbf{a}--\textbf{b},~Expectation value $\bra{0} I_{p,\sigma}(z) \ket{0}$ of the electric energy density along $z$ in the vacuum state $\ket{0}$ for (\textbf{a}) the $\sigma=x$ and (\textbf{b}) $\sigma=y$ polarizations and the mode $p=1$ (see Supplementary Section 4). The blue dashed line denotes the 2DEG location. \textbf{c}--\textbf{f},~Standard deviation $\sqrt{\bra{0} E^{2}_{p,\sigma}(\bm{\rho}) \ket{0}}$ of the in-plane electric field in the vacuum state (see Supplementary Section 4) at $z=z_\text{2DEG}$ for the $\sigma=x$ (\textbf{c}--\textbf{d}) and $\sigma=y$ (\textbf{e}--\textbf{f}) polarizations. The upper and lower panels correspond to the mode $p=1$ ($\omega_{1}/2\pi=338$\,GHz) and $p=2$ ($\omega_{2}/2\pi=382$\,GHz), respectively. The dashed rectangle at the center illustrates the silicon rod that is right above the 2DEG layer.}
\label{fig:asymmetric}
\end{figure}

The woodpile 3D-PCC exhibits 4 mirror symmetry planes per unit cell in the $xy$ plane. In the central unit cell, those mirror symmetry planes are $x=0;\pm a/2$ and $y=0;\pm a/2$ (see Fig.~\ref{fig:schematic}b). Depending on the incident polarization, $\sigma$, two families of cavity modes can be distinguished due to these symmetries. The modes excited by THz light polarized along $x$ are even (odd) with respect to $y=0;\pm a/2$ ($x=0;\pm a/2$). These modes are mostly linearly polarized in the $xz$-plane and are referred to as $\sigma=x$. On the other hand, the modes excited by THz light polarized along $y$ are odd (even) with respect to $y=0;\pm a/2$ ($x=0;\pm a/2$). These modes are mostly linearly polarized in the $yz$-plane and are referred to as $\sigma=y$. Furthermore, the spectrum of the 3D-PCC is degenerate for both $\sigma=x$ and $\sigma=y$ polarizations~\cite{VenturaGu2008OE} (see Supplementary Information Sec.~1). This is a consequence of the combination of a 4-fold rotational symmetry with respect to the $z$-axis and a mirror symmetry with respect to the plane $z=0$; The silicon rods above and below the GaAs defect layer are perpendicular to each other (Fig.~\ref{fig:schematic}b). For the cavity mode $p=1$, the expectation value of the electric energy density in the vacuum state $\ket{0}$, $\bra{0} I_{p,\sigma}(z) \ket{0}$, is maximum at the vertical location of the silicon rod that is adjacent to the defect layer and perpendicular to the polarization of incident THz light (Fig.~\ref{fig:asymmetric}a,b, Supplementary Information Sec.~4). Therefore, despite having the same resonance frequency, the cavity modes excited by orthogonal polarizations exhibit different spatial profiles. It should be noted that all cavity modes are localized in the vicinity of the defect GaAs layer with a standard deviation below one Si log thick for $p=3,4$ and about two logs thick for $p=1,2$ (see Supplementary Information Sec.~4).

To study the coupling between the cavity modes of the 3D-PCC and the CR of a 2DEG in GaAs, with dipole moment in the xy-plane, the GaAs defect layer is replaced by a GaAs substrate containing a multiple quantum well heterostructure near the top surface. The total thickness of the sample remains at around 60\,$\upmu$m. The standard deviation of the in-plane electric field in the EM vacuum state, $\sqrt{\bra{0} E^{2}_{p,\sigma}(\bm{\rho}) \ket{0}}$ (see Supplementary Information Sec.~4 for definition), for the first two bare cavity modes $p=1,2$ and $\sigma=x,y$ is displayed in Fig.~\ref{fig:asymmetric}c--f at the vertical location of the 2DEG, $z_\text{2DEG}$. For $\sigma=x$, the cavity electric field is tightly confined at the edges of the rods (Fig.~\ref{fig:asymmetric}c,d). Conversely, the mode profiles for $\sigma=y$ distribute more evenly within the unit cell (Fig.~\ref{fig:asymmetric}e,f). It should be noted that while the two families of modes $\sigma=x,y$ provide a good way to classify the cavity modes without 2DEG, those two polarizations are coupled by the off-diagonal elements of the dielectric tensor describing the CR (see Supplementary Information Sec.~5). However, it turns out that such a coupling is small and the polariton eigenmodes can thus still be well identified by their polarization index $\sigma$, as shown below.

\begin{figure}[ht!]%
\centering
\includegraphics[width=1.0\textwidth]{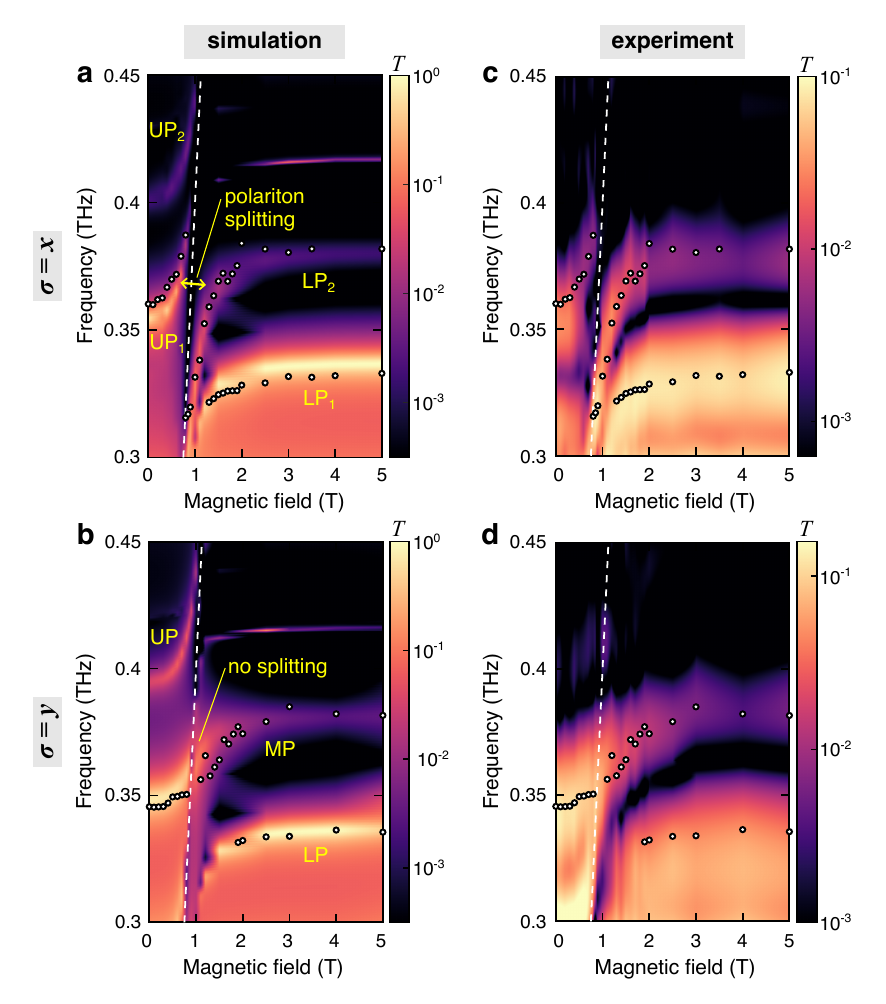}
\caption{\textbf{Polarization-dependent mixing of photonic modes.} \textbf{a}--\textbf{d},~Magnetic-field-dependent transmittance ($T$) spectra obtained from (\textbf{a}--\textbf{b}) numerical simulations and (\textbf{c}--\textbf{d}) experiments. The top (bottom) panels are for $\sigma=x$ ($\sigma=y$) polarizations. The white dots denote the peak frequencies extracted from experimental data using longer time-domain traces (Supplementary Information Sec.~8). The white dashed line shows the bare CR frequency. The LP--UP splitting is marked.}
\label{fig:polaritons}
\end{figure}

We conducted numerical simulations with COMSOL Multiphysics for the 3D-PCC with a GaAs 2DEG under an external magnetic field, $B$, along the $z$ direction (Supplementary Information Sec.~5). Figure~\ref{fig:polaritons}a,b shows transmittance spectra for the $x$ and $y$ polarizations, respectively. For simplicity, we restrict ourselves to the first two modes $p=1,2$ in the following discussion, since the modes $p=3,4$ with extremely narrow linewidths were not observed in the experimental data due to the limited resolution of THz spectroscopy. Four polariton branches, two upper polaritons (UPs) and two lower polaritons (LPs), are observed for the $x$ polarization (Fig.~\ref{fig:polaritons}a), accompanied by a splitting spanning the space around the line of the bare CR frequency (white dashed), $\omega_\text{c}$, that separates the UPs and LPs. By contrast, this splitting is not visible for the $y$ polarization (Fig.~\ref{fig:polaritons}b). Instead, we observed a S-shaped polariton branch that crosses the $\omega_\text{c}$ line and behaves as a LP for cavity mode $p=2$ at high $B$. The broad transmittance background at low frequencies is the photonic band edge, as shown in Fig.~\ref{fig:schematic}d.

We performed THz time-domain spectroscopy measurements to obtain the transmittance spectra under the same conditions. As the measurements directly collected the transmitted THz waveform in the time domain, the frequency resolution of the data is limited by the length of the time-domain traces that were used in the Fourier transformation. Note that the optical components in the system, for example, cryostat windows and crystals for THz generation and detection, induced echoes of the THz pulses at longer time delays. Including the echoes in the Fourier transformation causes artificial dips in the spectrum (Fabry--P\'erot effect)~\cite{LiEtAl2018NP}. Therefore, the time-domain traces used to plot the color plots in Fig.~\ref{fig:polaritons}c,d were truncated before the arrival of THz echoes (33\,ps) and zero padded to avoid the artificial dips. The frequency resolution of the color plot was limited. Furthermore, the white dots denote the $B$-dependent polariton frequencies that were extracted from longer time-domain traces through fittings. More details about the differences between short and long time-domain traces and the extraction procedures are discussed in the Supplementary Information Sec.~8. The experimental results show good agreement with the simulations, namely, we observed a polariton splitting that separates the UPs and LPs for the $x$ polarization and no splitting for the $y$ polarization. The UPs at high frequencies were not observed in experiments likely because the transmission is low for resonances near the center of the photonic band gap and also because there exist additional losses induced by imperfections in the fabricated woodpile lattice.

\subsection*{Theoretical analysis}
To gain a better understanding of the observed polarization-dependent transmittance spectra, we developed a microscopic quantum model described by the Hamiltonian $\hat{H} = \hat{H}_{\textrm{cav}} + \hat{H}_{\textrm{CR}} + \hat{H}_{\textrm{int}} + \hat{H}_{A^{2}}$ (Supplementary Section 2). The free-photon Hamiltonian is $\hat{H}_{\textrm{cav}} = \sum_{p,\sigma} \hbar \omega_{p} \hat{a}^{\dagger}_{p,\sigma} \hat{a}_{p,\sigma}$, with $\hat{a}_{p,\sigma}$ the annihilation operator of a photon in mode $p$ with polarization $\sigma$ and frequency $\omega_{p}$. The effective CR Hamiltonian $\hat{H}_{\textrm{CR}} = \hbar\omega_\text{c} \int \! \frac{d\bm{\rho}}{a} \hat{b}^{\dagger} (\bm{\rho}) \hat{b} (\bm{\rho})$ is written in terms of the collective CR operators at the in-plane position $\bm{\rho}$ in the woodpile unit cell. The associated creation operator, 
$$\hat{b}^{\dagger} (\bm{\rho})= \frac{1}{a\sqrt{\mathcal{N}}}\sum_{k,\textbf{G}} \hat{c}^{\dagger}_{\nu,k-G_{y}} \hat{c}_{\nu-1,k} e^{i G_{x} k l^{2}_\text{c}} e^{-i \textbf{G}\cdot \bm{\rho}},$$ promotes an electron from the highest-occupied Landau level (LL) $\nu-1$ with momentum $k$ to the lowest-unoccupied LL $\nu$ and momentum $k-G_{y}$, with $\hat{c}$ and $\hat{c}^{\dagger}$ the electron annihilation and creation operators, $G_{j}=m_{j}\times2\pi/a$ ($m_{j}=0,1,2,...$) the $j=x,y$ component of the in-plane reciprocal lattice vector of the 3D-PCC, $l_\text{c}=\sqrt{\hbar/eB}$ the magnetic length, $e$ the electron charge, $\nu$ the LL filling factor, and $\mathcal{N}$ the LL degeneracy (see Supplementary Information Sec.~2). The collective CR operators approximately satisfy bosonic commutation relations $[\hat{b} (\bm{\rho}),\hat{b}^{\dagger} (\bm{\rho}')]=\delta (\bm{\rho}-\bm{\rho}')$ in the dilute regime. We emphasize that those degenerate CR modes at each position $\bm{\rho}$ are extra degrees of freedom that are introduced in the effective Hamiltonian $\hat{H}_{\textrm{CR}}$ because they happen to be the modes coupled to the EM field (see below).


The next term in the Hamiltonian,
\begin{align}
\hat{H}_{\textrm{int}}&= i \hbar \sum_{p,\sigma} \int \! \frac{d\bm{\rho}}{a} g_{p,\sigma,x} (\bm{\rho}) \left[\hat{b}^{\dagger} (\bm{\rho}) - \hat{b} (\bm{\rho}) \right] \left(\hat{a}_{p,\sigma} + \hat{a}^{\dagger}_{p,\sigma}\right) \nonumber \\
&+ \hbar \sum_{p,\sigma} \int \! \frac{d\bm{\rho}}{a} g_{p,\sigma,y} (\bm{\rho}) \left[\hat{b}^{\dagger} (\bm{\rho}) + \hat{b} (\bm{\rho}) \right] \left(\hat{a}_{p,\sigma} + \hat{a}^{\dagger}_{p,\sigma}\right),
\label{H_int}
\end{align}
is the linear coupling between the CR and the EM field, whose strength is $g_{p,\sigma,j}(\bm{\rho}) = E_{p,\sigma,j}(\bm{\rho},z_\text{2DEG})\sqrt{e^{2} \omega_\text{c} n_\text{e}/(4 \varepsilon_{0} m_{\textrm{eff}} \omega_{p}a)}$. Here $j=x,y$ denotes the different components of the (real, dimensionless) electric field mode functions $E_{p,\sigma,j}(\bm{\rho},z_\text{2DEG})$ at the in-plane position $\bm{\rho}$ and vertical location of the QW, $m_{\textrm{eff}}$ is the electron mass in GaAs, $\varepsilon_{0}$ is the vacuum permittivity, and $n_\text{e}$ is the total electron density. The coupling of the CR excitations to each cavity mode $(p,\sigma)$ is weighted by the electric field mode functions since the electric field in the 3D-PCC is not uniform in the $xy$ plane, in contrast to 1D-PCCs~\cite{ZhangEtAl2016NP,LiEtAl2018NP}. Nevertheless, since the electric field mode functions are quasi-uniform over a cyclotron orbit, i.e. $a\ll l_\text{c}$, the dipole approximation can still be used to calculate the coupling matrix elements.

\begin{figure}[ht]%
\centering
\includegraphics[width=1\textwidth]{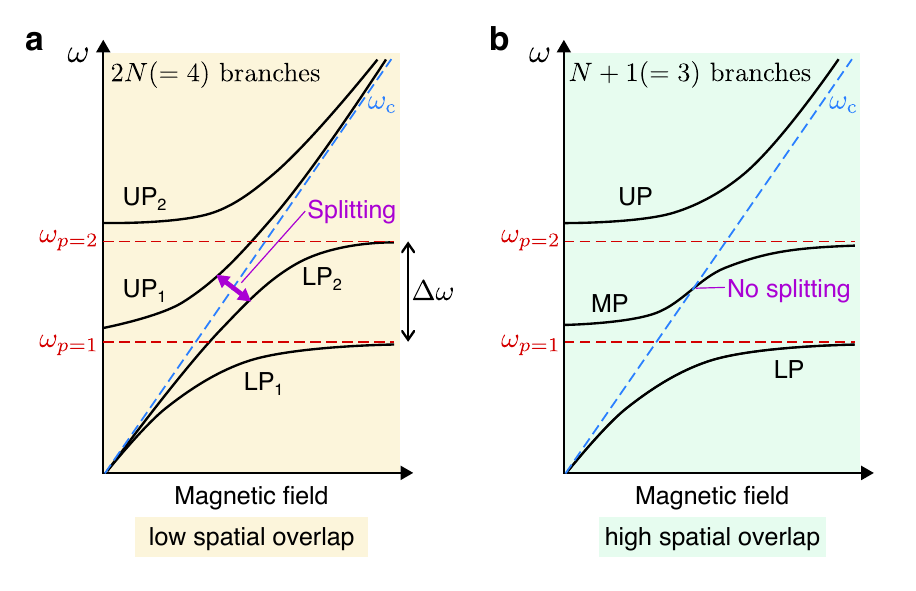}
\caption{\textbf{Different scenarios of multimode light--matter coupling.} \textbf{a}--\textbf{b},~Sketch of the typical polariton branches as a function of the magnetic field for two cavity modes ($N=2$, red dashed lines). The CR is depicted as a blue dashed line, $\omega_\text{c}$. When $g$ becomes larger than $\Delta\omega$ (superstrong coupling regime), there are two possible scenarios depending on the spatial overlap of the mode profiles: (\textbf{a}) the ``decoupled'' scenario with $2N$ polariton branches; and (\textbf{b}) the ``coupled'' scenario with $N+1$ polariton branches. In the latter case, the MP is a mixture of the two cavity modes close to the inflection point. UP: upper polariton; LP: lower polariton; MP: middle polariton.}
\label{fig:scenarios}
\end{figure}

The last term in the Hamiltonian, the so-called $A^{2}$ term,
\begin{align}
\hat{H}_{A^{2}} &= \sum_{p,p'} \sum_{\sigma,\sigma'} \hbar D_{p,p';\sigma,\sigma'}\left(\hat{a}_{p,\sigma} + \hat{a}^{\dagger}_{p,\sigma}\right) \left(\hat{a}_{p',\sigma'} + \hat{a}^{\dagger}_{p',\sigma'}\right),
\label{H_A2}
\end{align}
is responsible for direct coupling between the cavity modes. This term scales with $D_{p,p';\sigma,\sigma'}=\sum_{j}\int \! (d\bm{\rho}/a^{2}) g_{p,\sigma,j}(\bm{\rho})g_{p',\sigma',j}(\bm{\rho})/\omega_\text{c} \propto \sum_{j}\int \! d\bm{\rho} \, E_{p,\sigma,j}(\bm{\rho},z_\text{2DEG}) E_{p',\sigma',j}(\bm{\rho},z_\text{2DEG})$, and thus, involves an overlap integral between the in-plane spatial profiles of the modes. It should be emphasized that the cavity modes are normalized over the entire volume of the 3D-PCC, i.e., $\int \! d\bm{\rho} dz \, \varepsilon (\bm{\rho},z) {\bf E}_{p,\sigma} (\bm{\rho},z) \cdot {\bf E}_{p',\sigma'} (\bm{\rho},z) =a^{3} \delta_{p,p'}\delta_{\sigma,\sigma'}$, with $\varepsilon(\bm{\rho},z)$ the inhomogeneous dielectric profile. Thus, the cavity modes are orthogonal if the entire volume of the 3D-PCC is considered. Nevertheless, the in-plane overlap of the cavity mode profiles, which governs the mixing of the cavity modes mediated by the CR, is \textit{a priori} finite, as was recently reported in THz metamaterial resonators~\cite{CorteseEtAl2023O,MornhinwegEtAl2024N}. When the in-plane overlap is small, each cavity mode couples to the CR independently, and no matter-mediated intermode coupling exists. Each cavity mode leads to two polariton branches, resulting in a total of $2N$ branches~\cite{BalasubrahmaniyamEtAl2021PRB,CorteseEtAl2023O,MandalEtAl2023NL,GodsiEtAl2023TJoCP,MornhinwegEtAl2024N} (see Fig.~\ref{fig:scenarios}a for $N = 2$). By contrast, when the different cavity modes within the active region highly overlap, these modes can couple to each other through the matter, producing $N+1$ polariton branches with an S-shaped middle polariton (MP) (see Fig.~\ref{fig:scenarios}b for $N = 2$). Furthermore, when the light-matter coupling strength is comparable or larger than the frequency difference between the cavity modes, the system enters the superstrong coupling (SSC) regime~\cite{MeiserMeystre2006PRA,EggerWilhelm2013PRL,SundaresanEtAl2015PRX,KostylevEtAl2016APL,KuzminEtAl2019NQI,JohnsonEtAl2019PRL,MehtaEtAl2022}, and the mixing between different cavity modes increases, especially close to the inflection point of the S-shaped MP(Fig.~\ref{fig:scenarios}b).

To better understand how the two regimes with either mode coupling or mode decoupling emerge from the microscopic description, we introduce new CR excitation operators following the spatial profiles of the cavity modes, $\hat{b}_{p,\sigma}= \int \! (d\bm{\rho}/a) \, (\widetilde{g}_{p,\sigma}(\bm{\rho})/\Omega_{p,\sigma})  \hat{b}(\bm{\rho})$, with $\widetilde{g}_{p,\sigma}(\bm{\rho})=g_{p,\sigma,y}(\bm{\rho})-i g_{p,\sigma,x}(\bm{\rho})$ and the effective coupling strength $\Omega_{p,\sigma}$ defined by $\Omega^{2}_{p,\sigma} = \int \! (d\bm{\rho}/a^{2}) \, \vert \widetilde{g}_{p,\sigma}(\bm{\rho}) \vert^{2}$. By construction, the linear coupling term \eqref{H_int} written in terms of those operators takes the simple form 
\begin{align}
\hat{H}_{\textrm{int}}&=\sum_{p,\sigma} \hbar \Omega_{p,\sigma} \left(\hat{b}_{p,\sigma} + \hat{b}^{\dagger}_{p,\sigma} \right) \left(\hat{a}_{p,\sigma} + \hat{a}^{\dagger}_{p,\sigma}\right).
\label{Hintdec}
\end{align}
The commutation relations between the new CR modes read $[\hat{b}_{p,\sigma},\hat{b}^{\dagger}_{p',\sigma'}]= \xi_{p,p';\sigma,\sigma'}$, where 
\begin{align*}
\xi_{p,p';\sigma,\sigma'} = \frac{1}{\Omega_{p,\sigma} \Omega_{p',\sigma'}}\int \! \frac{d\bm{\rho}}{a^{2}} \, \widetilde{g}_{p,\sigma}(\bm{\rho}) \widetilde{g}^{*}_{p',\sigma'}(\bm{\rho}), \qquad (0 \leq \vert\xi_{p,p';\sigma,\sigma'}\vert\leq 1)
\end{align*}
is proportional to the in-plane spatial overlap of the different cavity modes, similarly as the off-diagonal contributions of the $A^{2}$ term. If the new CR modes $\hat{b}_{p,\sigma}$ were orthogonal, which is \textit{a priori} not the case, one would have $\xi_{p,p';\sigma,\sigma'} = \delta_{p,p'} \delta_{\sigma,\sigma'}$. The deviation of the parameter $\xi_{p,p';\sigma,\sigma'}$ with respect to $\delta_{p,p'} \delta_{\sigma,\sigma'}$, therefore, is a measure of how much coupling between the cavity modes is mediated by the CR.   

First, we find that $\xi_{p,p';x,y} < 0.36$ for all $p,p'$, indicating that the cavity modes with orthogonal polarizations are weakly coupled by the CR. Thus, the polarization index $\sigma$ remains quite a good quantum number to classify the polariton eigenmodes, as stated before.
For a given polarization $\sigma$, $\xi_{p,p';\sigma,\sigma} = 1$ ($\xi_{p,p';j,j} = 0$) corresponds to perfect overlap (no overlap) between the cavity modes $p$ and $p'$, in which case the cavity modes are coupled (decoupled) via the CR. For our 3D-PCC system, we find a rather small overlap between the spatial profiles of the $\sigma=x$ modes $p=1$ and $p=2$ ($\xi_{1,2;x,x} = 0.29$), suggesting that the intermode coupling is weak. Conversely, for $\sigma=y$ modes, the spatial profiles of cavity modes $p=1$ and $p=2$ highly overlap ($\xi_{1,2;y,y} = 0.91$), suggesting that they are strongly coupled through the CR. 

When $\xi_{p,p';\sigma,\sigma'} = \delta_{p,p'} \delta_{\sigma,\sigma'}$, and neglecting the coupling between different polarizations induced by the $A^{2}$ term, one has $D_{p,p';\sigma,\sigma} \approx (\Omega_{p,\sigma}\Omega_{p',\sigma}/\omega_{c})\xi_{p,p';\sigma,\sigma} = (\Omega^{2}_{p,\sigma}/\omega_{c}) \delta_{p,p'}$, and the full Hamiltonian can then be simplified to a ``decoupled'' Hamiltonian of the form $\hat{H} = \hat{H}_{\textrm{cav}} + \hat{H}_{\textrm{CR}} + \hat{H}_{\textrm{int}} + \hat{H}_{A^{2}}$, with $\hat{H}_{\textrm{cav}} = \sum_{p,\sigma} \hbar \omega_{p} \hat{a}^{\dagger}_{p,\sigma} \hat{a}_{p,\sigma}$, $\hat{H}_{\textrm{CR}} = \hbar\omega_\text{c} \sum_{p,\sigma}\hat{b}^{\dagger}_{p,\sigma} \hat{b}_{p,\sigma}$, $\hat{H}_{\textrm{int}}$ given by Eq.~\eqref{Hintdec}, and
\begin{align}
\hat{H}_{A^{2}} = \sum_{\sigma,p} \frac{\hbar\Omega^{2}_{p,\sigma}}{\omega_{c}} \left(\hat{a}_{p,\sigma} + \hat{a}^{\dagger}_{p,\sigma}\right)^{2}.
\label{Hfin_A3}
\end{align} 
In contrast to the full Hamiltonian, such a decoupled Hamiltonian can be diagonalized in each subspace $(p,\sigma)$ independently~\cite{MandalEtAl2023NL}.

\begin{figure}[ht!]%
\centering
\includegraphics[width=1\textwidth]{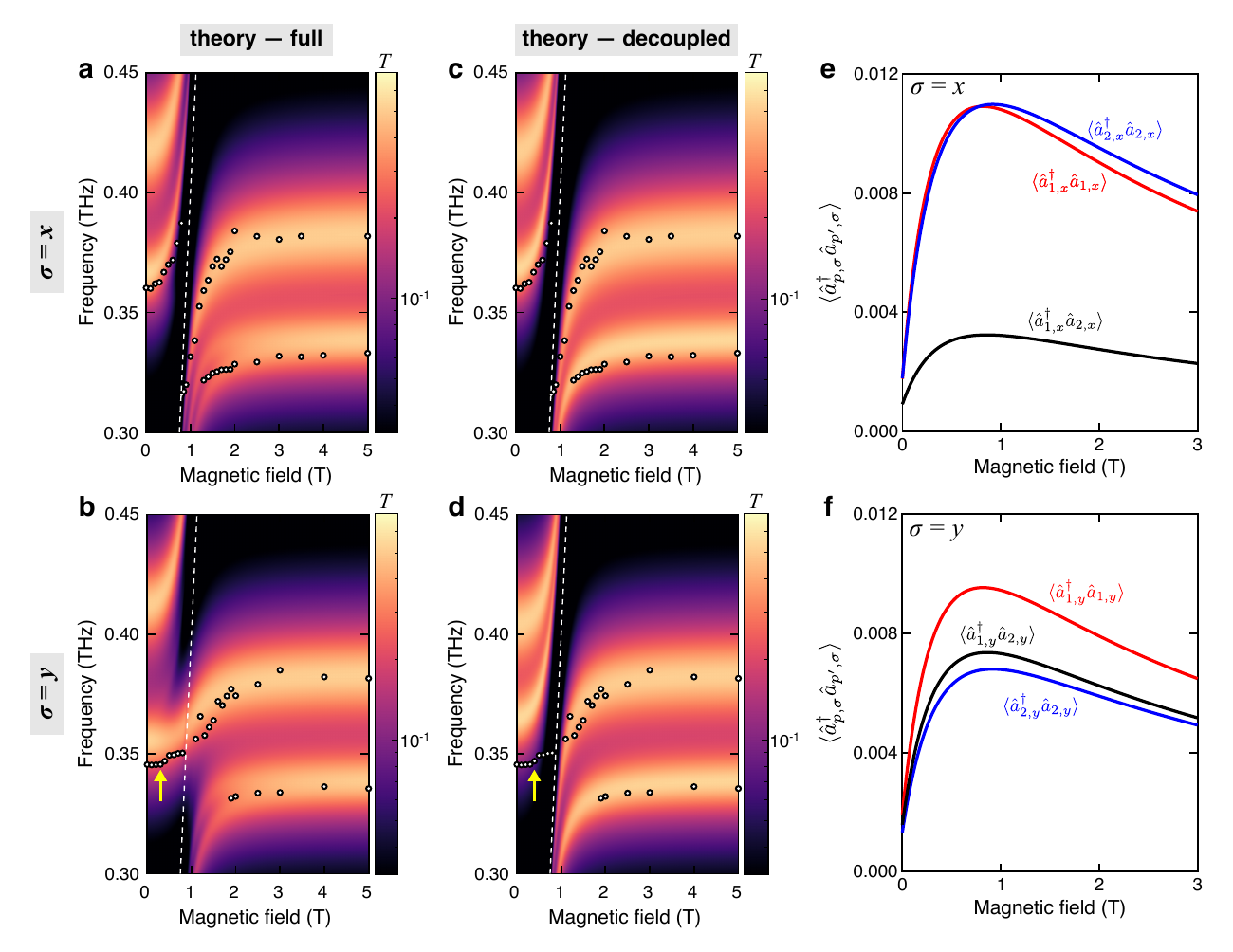}
\caption{\textbf{Extended Hopfield model for multimode light--matter coupling in the 3D-PCC.} \textbf{a}--\textbf{d},~Magnetic-field-dependent transmittance ($T$) spectra obtained from calculations using (\textbf{a}--\textbf{b}) the full Hamiltonian and (\textbf{c}--\textbf{d}) the decoupled Hamiltonian, respectively (Supplementary Information Sec.~2). The top (bottom) panels are for $\sigma=x$ ($\sigma=y$) modes. The white dots denote the peak frequencies extracted from experimental data using longer time-domain traces (Supplementary Information Sec.~8). The white dashed line shows the bare CR frequency. The yellow arrows in (\textbf{b},\textbf{d}) mark the UP that significantly deviates from the experimental data points in the calculation using the decoupled Hamiltonian. \textbf{e}--\textbf{f},~Calculated ground-state (virtual) correlations $\langle\hat{a}^{\dagger}_{p,\sigma}\hat{a}_{p',\sigma}\rangle$ using the full Hamiltonian for (\textbf{e}) $\sigma=x$ and (\textbf{f}) $\sigma=y$ modes. The ground-state correlations between the photonic modes $p=1$ and $p=2$ are significant for the $\sigma=y$ modes.}
\label{fig:theory}
\end{figure}

We computed transmittance spectra by extending the input--output model of Ref.~\cite{CiutiCarusotto2006PRA} in a simple planar geometry to a multimode PCC cavity (Supplementary Section 2). Dissipation is included through the (Markovian) coupling of cavity modes and CR excitations to phenomenological bosonic reservoirs, with associated quality factors of the bare cavity modes computed by FDTD, and of the intrinsic CR decay rate, respectively (Supplementary Information Sec.~2). The input--output model allows to select the polarization ($x,y$) of the input THz field to probe the polariton eigenmodes for both $\sigma=x$ and $\sigma=y$. The calculated spectra using the full Hamiltonian (Figs.~\ref{fig:theory}a and~\ref{fig:theory}b) are consistent with the simulations and experimental results (Fig.~\ref{fig:polaritons}), except for the transmittance background at low frequencies due to the lower edge of the photonic band gap that is not included in the model. In particular, the model replicates the existence of a LP--UP splitting close to the $\omega_c$ line and the emergence of an S-shaped MP that adiabatically transforms from the UP of mode 1 to the LP of mode 2 for $\sigma=x$ and $\sigma=y$ modes, respectively. Note that since the input field in the model is assumed to be independent of the cavity mode it couples to (Supplementary Information Sec.~2), the transmittance at high frequencies is overestimated as compared to the experimental data and numerical simulations. Since $\xi_{1,2;x,x}$ ($\sigma=x$) is small, the decoupled model provide a faithful description of the system, and thus the calculated spectra (Fig.~\ref{fig:theory}c) is almost the same as that using the full Hamiltonian (Fig.~\ref{fig:theory}a). For $\sigma=y$, however, the decoupled model is inaccurate as it fails to predict the emergence of the S-shaped MP (Fig.~\ref{fig:theory}d). This is due to the large overlap $\xi_{1,2;y,y}=0.91$, which means that the CR operators $\hat{b}_{1,y}$ and $\hat{b}^{\dagger}_{1,y}$ are not orthogonal, i.e., $[\hat{b}_{1,y},\hat{b}^{\dagger}_{2,y}] \neq 0$.

In order to assess in which regime of light-matter coupling our system operates, we extracted the coupling strengths for each cavity mode from the theoretical model and evaluated the relevant figures of merit. The USC regime, which has attracted widespread interest due to the compelling prospect of manipulating basic material properties by engineering the vacuum electromagnetic field surrounding the material inside a cavity~\cite{LiEtAl2018NP,NagarajanEtAl2020AN,AppuglieseEtAl2022S}, is achieved when the effective coupling strength for each cavity mode $\Omega_{p,\sigma}$ reaches about $10\%$ of the bare mode frequency at resonance. Here we find $\Omega_{1,x}/\omega_{1} \approx 0.21$ and $\Omega_{1,y}/\omega_{1} \approx 0.2$ at $B=0.81$T, and $\Omega_{2,x}/\omega_{2} \approx 0.21$ and $\Omega_{2,y}/\omega_{2} = 0.17$ at $B=0.92$T, respectively. This shows that our system operates in the USC regime for both cavity modes and both polarization $\sigma$.

It is interesting to investigate unique quantum features that can emerge in our multimode system. Since the ground state in the USC regime is known to be a dressed ground state that has a finite photon population, $\langle n_{p,\sigma}\rangle = \langle\hat{a}^{\dagger}_{p,\sigma}\hat{a}_{p,\sigma}\rangle$~\cite{CiutiEtAl2005PRB,CiutiCarusotto2006PRA}, an intriguing question is whether the virtual photons (cavity vacuum fields) in different modes can mix through simultaneous USC with matter, resulting in finite intermode correlations in the ground state $\langle\hat{a}^{\dagger}_{p,\sigma}\hat{a}_{p',\sigma}\rangle>0$. We computed the photon ground-state correlations $\langle\hat{a}^{\dagger}_{p,\sigma}\hat{a}_{p',\sigma}\rangle$ for both $\sigma=x,y$ by inverting the Bogoliubov transformation that diagonalizes the full (quadratic) Hamiltonian. The photon number for each mode $p$ and polarization $\sigma$ in the ground state (without external driving) is shown in Fig.~\ref{fig:theory}e. While $\langle\hat{a}^{\dagger}_{p,\sigma}\hat{a}_{p,\sigma}\rangle$ reaches its maximum at resonance between the CR and the cavity mode $(p,\sigma)$, the intermode correlation $\langle\hat{a}^{\dagger}_{1,\sigma}\hat{a}_{2,\sigma}\rangle$ is peaked in between the two resonances. For $\sigma=x$, we find that $\langle\hat{a}^{\dagger}_{1,x}\hat{a}_{2,x}\rangle \ll \langle\hat{a}^{\dagger}_{p,x}\hat{a}_{p,x}\rangle$. However, for $\sigma=y$, $\langle\hat{a}^{\dagger}_{1,y}\hat{a}_{2,y}\rangle$ is comparable to $\langle\hat{a}^{\dagger}_{p,y}\hat{a}_{p,y}\rangle$ (Fig.~\ref{fig:theory}f), which suggests that the mixing of virtual photons in different photonic modes is also directly related to the spatial overlap of the cavity profiles. Those intermode vacuum correlations should be accessible in our system from, e.g., first-order equal-time correlations measurements.  

Based on the scaling of the intermode ground-state correlations with the overlap between the cavity modes and the mode frequencies (see supplementary information Sec.~2), we extend the standard figure of merit for single mode USC to the multimode case by introducing the parameter
\begin{align}
\eta_{pp',\sigma} \equiv \sqrt{\frac{\int (d\bm{\rho}/a^{2}) \, \widetilde{g}_{p,\sigma}(\bm{\rho})\widetilde{g}^{*}_{p',\sigma}(\bm{\rho})}{\omega_\text{c}(\omega_{p}+\omega_{p'})/2}},
\end{align}
which does not depend on the magnetic field and whose diagonal elements ($p=p'$) coincide with the standard figure of merit. While intermode (off-diagonal) coupling for $\sigma=x$ is smaller than the diagonal coupling ($\eta_{12,x}=0.11$), we find that intermode coupling for $\sigma=y$ is of the same magnitude ($\eta_{12,y}=0.17$), consistently with the intermode ground-state correlations.

On the other hand, the SSC regime is achieved when the coupling strength becomes comparable to the frequency difference between the uncoupled modes, thereby signaling the breakdown of the single-mode approximation~\cite{MeiserMeystre2006PRA,EggerWilhelm2013PRL,SundaresanEtAl2015PRX,KostylevEtAl2016APL,KuzminEtAl2019NQI,JohnsonEtAl2019PRL,MehtaEtAl2022}. In this regime, hybridization of cavity mode profiles~\cite{MeiserMeystre2006PRA} and complex multimode dynamics~\cite{KrimerEtAl2014PRA,SundaresanEtAl2015PRX} have been discussed. However, our results highlight that such a criterion is not sufficient to characterize the SSC regime because the mixing between cavity modes is also influenced by the spatial profiles of the cavity modes. For example, two independent sets of LPs and UPs are observed for $\sigma=x$ as a result of the weak overlap of the cavity modes. Thus, the single-mode approximation still holds in this case even though the standard criterion is satisfied. As explained above, this ``mode decoupling'' situation would generally occur in all systems where quantum emitters fill the entire cavity mode volume~\cite{BalasubrahmaniyamEtAl2021PRB,CorteseEtAl2023O,MandalEtAl2023NL,GodsiEtAl2023TJoCP,MornhinwegEtAl2024N}. To circumvent this issue, we extend the standard figure of merit for the SSC regime to take into account the spatial overlap between the cavity modes,
\begin{align}
\Lambda_{\sigma} \equiv \sqrt{\frac{\int (d\bm{\rho}/a^{2}) \, \widetilde{g}_{1,\sigma}(\bm{\rho})\widetilde{g}^{*}_{2,\sigma}(\bm{\rho})}{\omega_\text{c}(\omega_{2}-\omega_{1})}}.
\end{align}
This parameter quantifies the strength of the intermode coupling with respect to the frequency difference $\omega_{2}-\omega_{1}$, including the effect of the cavity mode overlap. This parameter is $B$-independent as $\omega_c$ is included in the denominator. We find that $\Lambda_{x} \approx 0.33$ and $\Lambda_{y} \approx 0.49$, which indicates that the polarization $\sigma=y$ is further in the SSC regime than the polarization $\sigma=x$. Deep into the SSC regime ($\Lambda_{\sigma} \sim 1$), the S-shaped MP would become a pure-matter hybrid mode composed of the two cavity modes (see supplementary information Sec.~2). We show in Supplementary Information Sec.~2 that the SSC figure of merit governs the weight of the MP onto the different cavity modes, which in turn governs the intermode correlations in polaritonic excited states.

\section*{Discussion}\label{sec:discussion}
We reported ultrastrong and superstrong coupling in a THz 3D-PCC that is coupled to a Landau-quantized 2DEG in GaAs. The in-plane reciprocal lattice vector, $G$, or the spatial profiles of the cavity modes of the 3D-PCC affects the coupling between different cavity modes mediated by the matter. In contrast to usual planar cavities with mirror symmetry in the stacking direction, our cavity follows a combined symmetry of reflection across the $z=0$ plane and a 90$^\circ$ rotation. As a result, our system contains various degrees of hybridization between the cavity modes that can be selectively observed by rotating the polarization of the incident light while preserving USC between the CR and each cavity mode. Our results emphasize that the coupling strength of a multimode system cannot be accurately extracted from the vacuum Rabi splitting in the spectrum due to the formation of MPs when the cavity mode profiles highly overlap. Our experimental data showed good agreement with simulations and the calculations based on a microscopic model. We discussed the possible existence of quantum ground-state correlations between two photonic modes in the multimode USC regime that were derived from the theoretical model. We introduced relevant figures of merit to characterize the light-matter coupling regimes in multimode systems.

Although our system is linear in the sense that it is well described by a quadratic Hamiltonian, and thus does not feature photon-photon many-body interactions, the latter could be achieved by, e.g., replacing GaAs with a non-parabolic semiconductor or graphene, which would lead to a nonlinear CR. Introducing such nonlinearities in our system that operates in the USC and the SSC regimes simultaneously thus opens up interesting perspectives to explore both multimode vacuum-induced effects~\cite{HubenerEtAl2021NM,Garcia-VidalEtAl2021S,SchlawinEtAl2022APR,BlochEtAl2022N} and driven-dissipative dynamics in the many-body regime of quantum optics~\cite{EggerWilhelm2013PRL,SundaresanEtAl2015PRX,KuzminEtAl2019NQI,VrajitoareaEtAl2024}.

The 3D-PCC design is highly versatile and can be tailored to achieve photonic modes with smaller mode volumes and higher quality factors, for example, using a point defect~\cite{TaverneEtAl2015JOSAB}. Our approach can be utilized to investigate the coupling between the topological edge states of a 2DEG and the topological edge states of a chiral woodpile structure~\cite{ChangEtAl2017PRB,TakahashiEtAl2018JPSJ,VaidyaEtAl2020PRL,TakahashiEtAl2021OE,JorgEtAl2022LPR}. Moreover, the photonic modes with well-defined in-plane wavevectors in the 3D-PCC satisfy the condition to circumvent the no-go theorem for the Dicke SRPT~\cite{NatafEtAl2019PRL,AndolinaEtAl2020PRB,GuerciEtAl2020PRL,ManzanaresEtAl2022PRB}. While the photon wavevector $\sim2\pi/a$ is limited by the lattice parameter $a$ of the 3D-PCC and remains relatively small compared to the electron momentum $\sim1/l_\text{c}$, the possibility of fulfilling the criterion for 2D photon condensation in this regime has been recently discussed~\cite{AndolinaEtAl2020PRB}. Future work can focus on the predicted renormalization of matter's kinetic energy and effective mass at finite in-plane wave vectors near the photonic band edge~\cite{YangJohn2007PRB}.

\backmatter

\bmhead*{Supplementary information.}
Supplementary information is available for this paper at here.

\bmhead*{Acknowledgments.}
This work was done in part using resources of the Research Support Shop and Shared Equipment Authority at Rice University, as well as computational resources of the Centre de calcul de l'universit\'{e} de Strasbourg (CCUS). We thank Motoaki Bamba, Kaden Hazzard, Tal Schwartz, Thibault Chervy, and Cyriaque Genet for useful discussions.

\section*{Declarations}
\begin{itemize}
\item Funding
\begin{description}
\item J.K.\ acknowledges support from the U.S.\ Army Research Office (through Award No.\ W911NF2110157), the Gordon and Betty Moore Foundation (through Grant No.\ 11520), the W.\ M.\ Keck Foundation (through Award No.\ 995764), and the Robert A.\ Welch Foundation (through Grant No.\ C-1509).
\end{description}
\item Conflict of interest/Competing interests 
\begin{description}
\item The authors declare no competing interests.
\end{description}
\item Ethics approval 
\begin{description}
\item Not applicable.
\end{description}
\item Consent to participate
\begin{description}
\item Not applicable.
\end{description}
\item Consent for publication
\begin{description}
\item Not applicable.
\end{description}
\item Data availability
\begin{description}
\item The experimental data and numerical simulation that support the plots in this paper are available from the corresponding author upon reasonable request.
\end{description}
\item Code availability 
\begin{description}
\item The codes used in the theory part of this study are available from the corresponding author upon reasonable request.
\end{description}
\item Authors' contributions
\begin{description}
\item F.T., D.H., and J.K.\ conceptualized the project. F.T. and A.M.\ designed and fabricated the cavity device. F.T., A.B., and H.X.\ performed the measurements. F.T.\ analyzed the experimental data. D.H.\ derived the microscopic model and performed MEEP simulations and calculations. S.L., G.C.G., and M.J.M.\ grew the 2DEG sample by the molecular beam epitaxy system. F.T., S.S., and A.A.\ conducted COMSOL simulations. A.B., D.H., and J.K.\ supervised the project. F.T., D.H., and J.K.\ wrote the manuscript, with inputs from all authors.
\end{description}
\end{itemize}

\section*{Methods}\label{sec11}
\subsection*{Preparation of the multiple quantum well sample}
A wafer containing multiple GaAs quantum wells (MQW) was grown by the Purdue molecular beam epitaxy system. This structure had ten 30-nm-thick GaAs MQW separated by 160-nm Al$_{0.24}$Ga$_{0.76}$As barriers. Silicon dopants were placed 80\,nm away from the GaAs MQW. An electron density per well of 3.08~$\times$~10$^{11}$~cm$^{-2}$ and a mobility of 2.36~$\times$~10$^7$~cm$^2$/Vs were extracted from Hall transport measurements at 300\,mK in the dark. The total electron density of the MQW was 3.08~$\times$~10$^{12}$~cm$^{-2}$.

\subsection*{Fabrication of the THz woodpile cavity}
A 100-$\upmu$m-thick silicon wafer was coated by lift-off resist and photoresist layers through spin coating. The layers were patterned by photolithography. An $\sim$\,1.2-$\upmu$m-thick Al$_2$O$_3$ layer was deposited on the coated silicon wafer by an e-beam evaporator. The lift-off process was completed by removing the resist with Remover PG; leaving Al$_2$O$_3$ on the patterned area. Afterwards, the silicon wafer was etched by deep reactive ion etching using a Bosch process. During this process, the silicon wafer was cut into multiple smaller pieces, and the gaps between silicon rods and holes at the corners were also created in each piece of the smaller wafer. The remaining Al$_2$O$_3$ layer was removed by immersing the sample in a heated 1:3 solution of phosphoric acid and sulphuric acid. The width of the silicon rods and the periodicity were 87\,$\upmu$m and 333\,$\upmu$m, respectively. A custom-made sample holder was used for stacking the silicon wafers to form a woodpile cavity. The orientation of the silicon rods was fixed by the screws that passed through the holes at the corners. The woodpile cavity can be conveniently disassembled and reassembled. 

\subsection*{THz time-domain magnetospectroscopy measurements}
Transmittance spectra for samples were measured by a home-built THz time-domain magnetospectroscopy setup~\cite{WangEtAl2007OL,WangEtAl2010OE,ArikawaEtAl2011PRB,ArikawaEtAl2012OE,ZhangEtAl2014PRL,ZhangEtAl2016NP,LiEtAl2018NP,LiEtAl2018S,LiEtAl2019PRB,BaydinEtAl2020FO,BaydinEtAl2022PRL}. The near-infrared (775\,nm) output beam of a Ti:sapphire regenerative amplifier (1\,kHz, 200\,fs, Clark-MXR, Inc., CPA-2001) was split into pump and probe beams. The pump beam was used to generate linearly polarized THz radiation in a ZnTe crystal through optical rectification. The generated THz beam was focused onto the sample that was mounted on the sample holder of a 10-T superconducting magnet cryostat (Oxford Instruments, Inc.). The direction of the static magnetic field was normal to the sample surface. The time-domain waveform of the transmitted THz radiation, $E_\text{sample}(t)$, was measured with another ZnTe crystal via electro-optic sampling with controlled time delays. A reference signal, $E_\text{reference}(t)$, was measured by repeating the measurements in the absence of a sample. $E_{\text{sample}}(t)$ and $E_{\text{reference}}(t)$ were Fourier-transformed into complex-valued frequency-domain spectra, $\tilde{E}_{\text{sample}}(\omega)$ and $\tilde{E}_{\text{reference}}(\omega)$, respectively. Transmittance, $T$, is equal to $|\tilde{E}_{\text{sample}}(\omega)|^2/|\tilde{E}_\text{reference}(\omega)|^2$. The extraction of peak frequencies is discussed in Supplementary section 7 and 8.

\bibliography{USC}%

\end{document}


\title[Article Title]{Supplementary Information: Multimode ultrastrong coupling in three-dimensional photonic-crystal cavities}

\author[1,2]{\fnm{Fuyang} \sur{Tay}}
\author[1]{\fnm{Ali} \sur{Mojibpour}}
\author[1]{\fnm{Stephen} \sur{Sanders}}
\author[3,4]{\fnm{Shuang} \sur{Liang}}
\author[5]{\fnm{Hongjing} \sur{Xu}}
\author[6]{\fnm{Geoff} \sur{C. Gardner}}
\author[1,7]{\fnm{Andrey} \sur{Baydin}}
\author[3,4,6,8]{\fnm{Michael} \sur{J. Manfra}}
\author[1]{\fnm{Alessandro} \sur{Alabastri}}
\author[9]{\fnm{David} \sur{Hagenm\"uller}}
\author*[1,5,7,10]{\fnm{Junichiro} \sur{Kono}}\email{kono@rice.edu}

\affil[1]{\orgdiv{Department of Electrical and Computer Engineering}, \orgname{Rice University}, \orgaddress{\city{Houston}, \postcode{77005}, \state{Texas}, \country{USA}}}
\affil[2]{\orgdiv{Applied Physics Graduate Program, Smalley--Curl Institute}, \orgname{Rice University}, \orgaddress{\city{Houston}, \postcode{77005}, \state{Texas}, \country{USA}}}
\affil[3]{\orgdiv{Department of Physics and Astronomy}, \orgname{Purdue University}, \orgaddress{\city{West Lafayette}, \postcode{47907}, \state{Indiana}, \country{USA}}}
\affil[4]{\orgdiv{Birck Nanotechnology Center}, \orgname{Purdue University}, \orgaddress{\city{West Lafayette}, \postcode{47907}, \state{Indiana}, \country{USA}}}
\affil[5]{\orgdiv{Department of Physics and Astronomy}, \orgname{Rice University}, \orgaddress{\city{Houston}, \postcode{77005}, \state{Texas}, \country{USA}}}
\affil[6]{\orgdiv{School of Electrical and Computer Engineering}, \orgname{Purdue University}, \orgaddress{\city{West Lafayette}, \postcode{47907}, \state{Indiana}, \country{USA}}}
\affil[7]{\orgdiv{Smalley--Curl Institute}, \orgname{Rice University}, \orgaddress{\city{Houston}, \postcode{77005}, \state{Texas}, \country{USA}}}
\affil[8]{\orgdiv{School of Materials Engineering}, \orgname{Purdue University}, \orgaddress{\city{West Lafayette}, \postcode{47907}, \state{Indiana}, \country{USA}}}
\affil[9]{\orgdiv{CESQ-ISIS (UMR 7006)}, \orgname{Universit\'e de Strasbourg and CNRS}, \orgaddress{\city{Strasbourg}, \postcode{67000}, \country{France}}}
\affil[10]{\orgdiv{Department of Materials Science and NanoEngineering}, \orgname{Rice University}, \orgaddress{\city{Houston}, \postcode{77005}, \state{Texas}, \country{USA}}}

\newpage

\maketitle
\tableofcontents

\renewcommand{\figurename}{Fig.}
\renewcommand{\thefigure}{S\arabic{figure}}
\renewcommand{\theequation}{S\arabic{equation}}

\section{Transmittance spectrum for the bare 3D-PCC}
\label{sec:transmittance spectrum}

\begin{figure}[ht!]%
\centering
\includegraphics[width=0.85\textwidth]{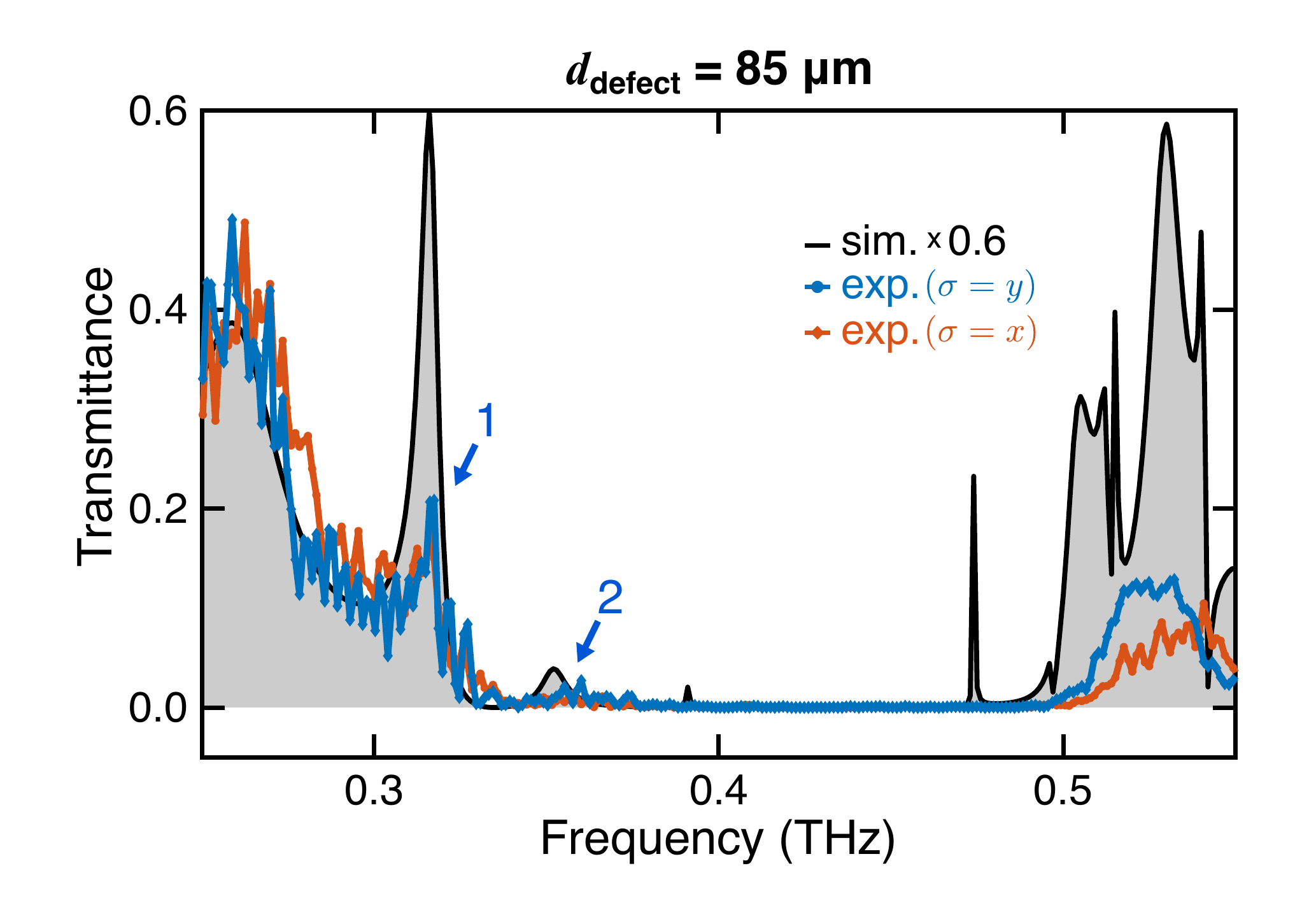}
\caption{\textbf{Transmittance spectrum for the bare 3D-PCC with a 85-$\upmu$m-thick GaAs layer.} We performed transmission measurements on a bare 3D-PCC, i.e., the defect layer is a bare GaAs substrate without the 2DEG layer. The black, blue, and red lines denote the spectra obtained from numerical simulation and experiments with $\sigma=y$- and $\sigma=x$-polarized incident THz radiation, respectively. The blue arrows mark cavity modes 1 and 2.}
\label{fig:bare3DPCC}
\end{figure}

We performed transmission measurements on a bare 3D-PCC, i.e., a bare GaAs substrate without the 2DEG layer, to examine the quality of the fabricated 3D-PCC, see Fig.~\ref{fig:bare3DPCC}. The GaAs substrate was polished down to $\sim$85\,$\upmu$m due to the inaccuracy of the polishing process. Although the bare GaAs substrate has a thickness that was slightly different from the actual GaAs sample with the 2DEG layer, the thickness discrepancy only caused shifts of photonic mode frequencies. The frequencies of modes 1 and 2 in the experimental data are consistent with the simulation with an 85-$\upmu$m-thick GaAs layer. The measured transmittance spectrum confirmed that the cavity modes of the 3D-PCC are degenerate for the two polarizations $\sigma=x$ and $\sigma=y$. 

\section{Microscopic model}
\label{sec:microscopic model}
The microscopic Hamiltonian is derived in the Coulomb gauge by extending the models of Refs.~\cite{HagenmullerEtAl2010PRB,Hagenmuller2016PRB,LiEtAl2018NP}. We use the vector potential operator
\begin{align}
\hat{\textbf{A}} (\bm{\rho},z)=\sum_{p,\sigma} \sqrt{\frac{\hbar}{2\varepsilon_{0}\omega_{p}a^{3}}} \, \textbf{E}_{p,\sigma} (\bm{\rho},z) \left(\hat{a}_{p,\sigma} + \hat{a}^{\dagger}_{p,\sigma} \right).
\end{align}
At normal incidence ($\Gamma$ point of the first Brillouin zone), the electric field mode functions $\textbf{E}_{p,\sigma} (\bm{\rho},z)$ can be chosen real without loss of generality. The calculation can be trivially extended to the full Brillouin zone using complex fields with a quasi-momentum $\textbf{q}$ as computed by FDTD. The linear coupling term between light and matter reads
\begin{align}
\frac{e}{m_{\textrm{eff}}}\int \! d\bm{\rho} dz \, \hat{\Psi}^{\dagger} (\bm{\rho},z) \hat{{\bm \pi}} \cdot \hat{\textbf{A}} (\bm{\rho},z) \hat{\Psi} (\bm{\rho},z),
\label{matrel2}
\end{align}
with $\hat{{\bm \pi}}$ the gauge-invariant in-plane momentum including the contribution of the static magnetic field $B$, and the fermion field $\hat{\Psi} (\bm{\rho},z) = L^{-1/2} \sum_{n,k} e^{-i k y} \chi_{n,k} (x) \xi (z) \hat{c}_{n,k}$~\cite{HagenmullerEtAl2010PRB}. We use the Landau gauge, where single electron states are characterized by an integer $n$ and the $y$ component of the electron momentum $k$. The wave functions are plane waves in the $y$ direction ($L$ is the length of the 2DEG in the plane), and those of a harmonic oscillator (Hermite polynomial of order $n$) centered at the guiding center position $k\, l^{2}_\text{c}$, with $l_\text{c}=\sqrt{\hbar/eB}$ the magnetic length. $\xi (z)$ is a normalized wave function that describes the QW confinement in the $z$ direction. $c^{\dagger}_{n,k}$ and $c_{n,k}$ are the creation and annihilation operators of an electron in the state $(n,k)$.

When computing the matrix elements entering Eq.~\eqref{matrel2}, we assume that mode spatial profiles $\textbf{E}_{p,\sigma} (\bm{\rho},z)$ remains constant over the QW thickness, and exploit the discrete translational invariance in the plane by decomposing the mode profiles into Fourier series $\textbf{E}_{p,\sigma} (\bm{\rho},z_\text{2DEG}) = \sum_{\textbf{G}} \textbf{U}_{p,\sigma}(\textbf{G}) e^{i \textbf{G}\cdot \bm{\rho}}$, with $\textbf{G}= (2\pi m_{x}/a) \textbf{e}_{x} + (2\pi m_{y}/a) \textbf{e}_{y}$ ($m_{x},m_{y} \in \mathbb{N}$) the reciprocal lattice vectors. One is then left with the calculation of overlap integrals of the kind
\begin{align}
I^{n'\pm 1,k'}_{n,k} (\textbf{G})=\int \! \frac{dy}{L} \, e^{i(k-k'+ G_{y})y} \int \! dx \, \chi_{n,k} (x) \chi_{n'\pm 1,k'} (x) e^{i G_{x} x}.
\label{overinteg2}
\end{align}
While the integral in the $y$ direction simply provides the selection rule $k'=k+G_{y}$, the calculation of the integral along $x$ is performed using the dipole approximation. The mode functions $\textbf{E}_{p,\sigma}(\bm{\rho},z_\text{2DEG})$ are slowly varying over the typical extent of the harmonic oscillator wave functions $\chi_{n,k} (x)$. For a magnetic field $B\simeq 1$\,T, the latter is indeed $\sim l_\text{c}\simeq 10-100\,{\rm nm}$, which is much smaller than the woodpile lattice parameter $a=333\,\upmu$m. Since the Fourier coefficients $\textbf{U}_{p,\sigma} (\textbf{G})$ exhibit large peaks at $G_{j} \sim 1/a$ ($m_{x} \sim m_{y} \sim 1$) and rapidly decrease as $G_{j} a \to \infty$, one has $G_{j} l_\text{c} \sim l_\text{c}/a \ll 1$. Equation \eqref{overinteg2} thus provides $I^{n'\pm 1,k'}_{n,k} (\textbf{G}) \approx \delta_{k',k+G_{y}}\delta_{n,n'\pm 1} e^{i G_{x} k l^{2}_\text{c}}$.

We introduce the CR excitation creation operator $$\hat{b}^{\dagger} (\bm{\rho})= \frac{1}{a\sqrt{\mathcal{N}}}\sum_{k,\textbf{G}} \hat{c}^{\dagger}_{\nu,k-G_{y}} \hat{c}_{\nu-1,k} e^{i G_{x} k l^{2}_\text{c}} e^{-i \textbf{G}\cdot \bm{\rho}},$$ which promotes an electron with momentum $k$ in the highest-occupied Landau level (LL) $n=\nu-1$ to the lowest-unoccupied LL $n=\nu$ and momentum $k-G_{y}$. $\nu=2\pi n_\text{e}l^{2}_\text{c}$ denotes the filling factor of the 2DEG and $\mathcal{N}$ is the LL degeneracy. The in-plane position vector $\bm{\rho}$ is restricted to a woodpile unit cell. The CR excitation operators $\hat{b} (\bm{\rho})$ and $\hat{b}^{\dagger} (\bm{\rho})$ satisfy bosonic commutation relations when the number of CR excitations remains small compared to $\mathcal{N}$, i.e., $\langle [\hat{b} (\bm{\rho}),\hat{b}^{\dagger} (\bm{\rho}')] \rangle=\delta (\bm{\rho}-\bm{\rho}')$, where $\langle \cdots \rangle$ denotes the expectation value in the electronic ground state (in which the lowest $\nu$ LLs are fully occupied). Note that the LL degeneracy entering the definition of the CR excitation operators is here written as $\mathcal{N} = a^{2}/(2\pi l^{2}_\text{c})$, using the lattice constant $a$ instead of the length $L$ of the 2DEG. This is indeed the only physical choice as the light--matter coupling strength should not depend explicitly on any length scale other than $l_\text{c}$ in the plane. With those definitions, one recovers the light-matter coupling term $H_{\textrm{int}}$ discussed in the main text. The $A^{2}$ term is derived in a similar fashion.

The transmission spectra shown in the main text were computed using an input--output model, which is an extension of the one introduced in Ref.~\cite{CiutiCarusotto2006PRA} in a simple planar geometry. Transmission of THz radiation through the 3D-PCC is modeled by introducing two identical photon reservoirs on each side of the cavity in the $z$ direction (top, bottom). Similarly, CR excitations of the 2DEG acquire a finite lifetime by interacting with a phenomenological bosonic reservoir. The total Hamiltonian includes the contribution (neglecting counter-rotating terms)
\begin{align}
\hat{H}_{\textrm{R}}&=\sum_{p,\sigma,\lambda} \int \! dq \, w_{p,\sigma} (q) \hat{\alpha}^{\dagger}_{p,\sigma,\lambda} (q) \hat{\alpha}_{p,\sigma,\lambda} (q) + i  \kappa_{p,\sigma} (q) \left[\hat{\alpha}_{p,\sigma,\lambda} (q) \hat{a}^{\dagger}_{p,\sigma} - \hat{a}_{p,\sigma} \hat{\alpha}^{\dagger}_{p,\sigma,\lambda} (q) \right] \nonumber \\
&+\int \! dq \int \! d\bm{\rho} \, \widetilde{w} (q,\bm{\rho}) \hat{\beta}^{\dagger} (\bm{\rho},q) \hat{\beta} (\bm{\rho},q) + i \widetilde{\kappa} (\bm{\rho},q) \left[\hat{\beta} (\bm{\rho},q) \hat{b}^{\dagger} (\bm{\rho}) - \hat{b} (\bm{\rho}) \hat{\beta}^{\dagger} (\bm{\rho},q) \right].
\label{Hres2}
\end{align}
The first contributions in the first and second lines of Eq.~\eqref{Hres2} describe the energy of the photonic modes in each reservoir $\lambda=(\textrm{top},\textrm{bot})$ and that of the matter excitation reservoir, respectively. $q$ denotes a phenomenological (continuous) parameter ensuring that the reservoirs have continuous spectra. The second terms in both lines describe the coupling between the mode of the reservoirs and the modes of the system, which provides the latter with a finite lifetime. The input--output method consists in solving the equations of motion of the system by introducing ``input'' and ``output'' operators: $\hat{\alpha}^{\textrm{in}}_{p,\sigma,\lambda} (q)=\lim_{t \to -\infty} \hat{\alpha}_{p,\sigma,\lambda} (q,t) e^{i w_{p,\sigma} (q) t}$, $\hat{\alpha}^{\textrm{out}}_{p,\sigma,\lambda} (q)=\lim_{t \to +\infty} \hat{\alpha}_{p,\sigma,\lambda} (q,t) e^{i w_{p,\sigma}(q) t}$, and similar expressions for the matter excitation reservoir operators. One can then find a linear relation between the input and output operators, which allows to compute the normalized transmission spectrum (transmitted light with polarization $\sigma'$, incident light with polarization $\sigma$) as
\begin{align*}
T_{\sigma,\sigma'}(\omega)=\frac{\sum_{p'} \langle \hat{\alpha}^{\textrm{out}\dagger}_{p',\sigma',\textrm{bot}} (q) \hat{\alpha}^{\textrm{out}}_{p',\sigma',\textrm{bot}} (q) \rangle}{\sum_{p} \langle \hat{\alpha}^{\textrm{in}\dagger}_{p,\sigma,\textrm{top}} (q) \hat{\alpha}^{\textrm{in}}_{p,\sigma,\textrm{top}} (q) \rangle}.
\end{align*}
We assume THz radiation coming from the top reservoir and populating all photonic modes equally, i.e., $\langle \hat{\alpha}^{\textrm{in}\dagger}_{p,\sigma,\textrm{top}} (q) \hat{\alpha}^{\textrm{in}}_{p,\sigma,\textrm{top}} (q) \rangle$ does not depend on $p$ nor $\sigma$, while $\langle \hat{\alpha}^{\textrm{in}\dagger}_{p,\sigma,\textrm{bot}} (q) \hat{\alpha}^{\textrm{in}}_{p,\sigma,\textrm{bot}} (q) \rangle=0 \; \forall \, (p,\sigma)$ and $\langle \hat{\beta}^{\textrm{in}\dagger} (\bm{\rho},q)\hat{\beta}^{\textrm{in}} (\bm{\rho},q) \rangle=0 \; \forall \, \bm{\rho}$. The output expectation value $\langle \hat{\alpha}^{\textrm{out}\dagger}_{p,\sigma,\textrm{bot}} (q) \hat{\alpha}^{\textrm{out}}_{p,\sigma,\textrm{bot}} (q) \rangle$ depends on the decay rates of each photonic modes $\Gamma_{p,\sigma}=\omega_{p}/Q_{p,\sigma}\equiv\pi \kappa^{2}_{p,\sigma} (q_{0}) \rho (\omega)$, with $Q_{p,\sigma}$ the quality factor of the cavity mode $(p,\sigma)$ as computed in FDTD (MEEP implementation~\cite{OskooiEtAl2010CPC}) without the 2DEG, and on the CR excitation decay rate $\Gamma_\text{c} \equiv \pi \widetilde{\kappa}^{2} (\bm{\rho},\widetilde{q}_{0}) \rho (\omega)$. The latter corresponds to the intrinsic CR decay rate. The quality factors of the photonic modes were obtained by fitting the peaks in transmittance spectra with Lorentzian functions. However, deviations from pure Lorentzian line shapes were clearly observed (especially for low-$Q$ modes), similar to a previous report for 0D-PCCs~\cite{PellegrinoEtAl2020PRL}, which has been attributed to dissipation (radiative leakage of the cavity modes)~\cite{SauvanEtAl2013PRL}. Note that the CR decay rate is suppressed when the 2DEG is placed inside a high-$Q$ cavity~\cite{ZhangEtAl2016NP}. As a good estimate, we used the CR decay rate obtained from previous experiments with a 1D-PCC ($\Gamma_\text{c}/2\pi = 5.7$\,GHz)~\cite{ZhangEtAl2016NP}. In the previous equations, $q_{0}$ and $\widetilde{q}_{0}$ are solutions of $\omega - w_{p,\sigma} (q)=0$ and $\omega-\widetilde{w} (q,\bm{\rho})=0$, respectively, and $\rho(\omega)$ is the effective density of states of the reservoirs, which are assumed to be Markovian, i.e., $\rho(\omega)$ is supposed to be frequency-independent in the range of interest. 

In order to further characterize the full and decoupled models discussed in the main text, we introduce a toy model with two cavity modes and one component for the electric field, one polarization, and continuously tunable spatial overlap between the cavity modes. The toy model Hamiltonian reads
\begin{align}
H&=\sum_{p} \hbar \omega_{p} \hat{a}^{\dagger}_{p} \hat{a}_{p} + \hbar\omega_\text{c} \int \! d\bm{\rho} \, \hat{b}^{\dagger} (\bm{\rho}) \hat{b} (\bm{\rho}) \nonumber \\
&+ \sum_{p} \int \! \frac{d\bm{\rho}}{a} \hbar g_{p} (\bm{\rho}) \left[\hat{b} (\bm{\rho}) + \hat{b}^{\dagger} (\bm{\rho}) \right] \left(\hat{a}_{p} + \hat{a}^{\dagger}_{p}\right) \nonumber \\
& + \sum_{p,p'} \int \! \frac{d\bm{\rho}}{a^{2}} \, \frac{\hbar g_{p}(\bm{\rho})g_{p'}(\bm{\rho})}{\omega_\text{c}} \left(\hat{a}_{p} + \hat{a}^{\dagger}_{p}\right) \left(\hat{a}^{\dagger}_{p'} + \hat{a}_{p'}\right),
\label{Hfin_A2_toy2}
\end{align}
with the coupling strength $g_{p}(\bm{\rho})=E_{p}(\bm{\rho})\sqrt{e^{2} \omega_\text{c} n_\text{e}/(4 \varepsilon_{0} m_{\textrm{eff}}\omega_{p}a)}$, and the in-plane mode profiles
\begin{align}
E_{1}(\bm{\rho}) &= \sin \left( \frac{2\pi x}{a} \right)\sin \left( \frac{2\pi y}{a} \right) \nonumber \\
E_{2}(\bm{\rho}) &= \sin \left( \frac{2\pi x}{a} +\frac{(1-\epsilon)\pi}{2} \right) \sin \left( \frac{2\pi y}{a} +\frac{(1-\epsilon)\pi}{2} \right). 
\label{mode_prog}
\end{align}
The parameter $\epsilon \in [0,1]$ allows us to artificially tune the spatial overlap between the two cavity modes. As $\epsilon$ is increased from $0$ (no overlap) to $1$ (perfect overlap), the splitting between the UP of the first mode $p=1$ and the LP of the second one $p=2$ becomes narrower and vanishes at $\epsilon=1$. 

\begin{figure}[ht!]%
\centering
\includegraphics[width=0.9\textwidth]{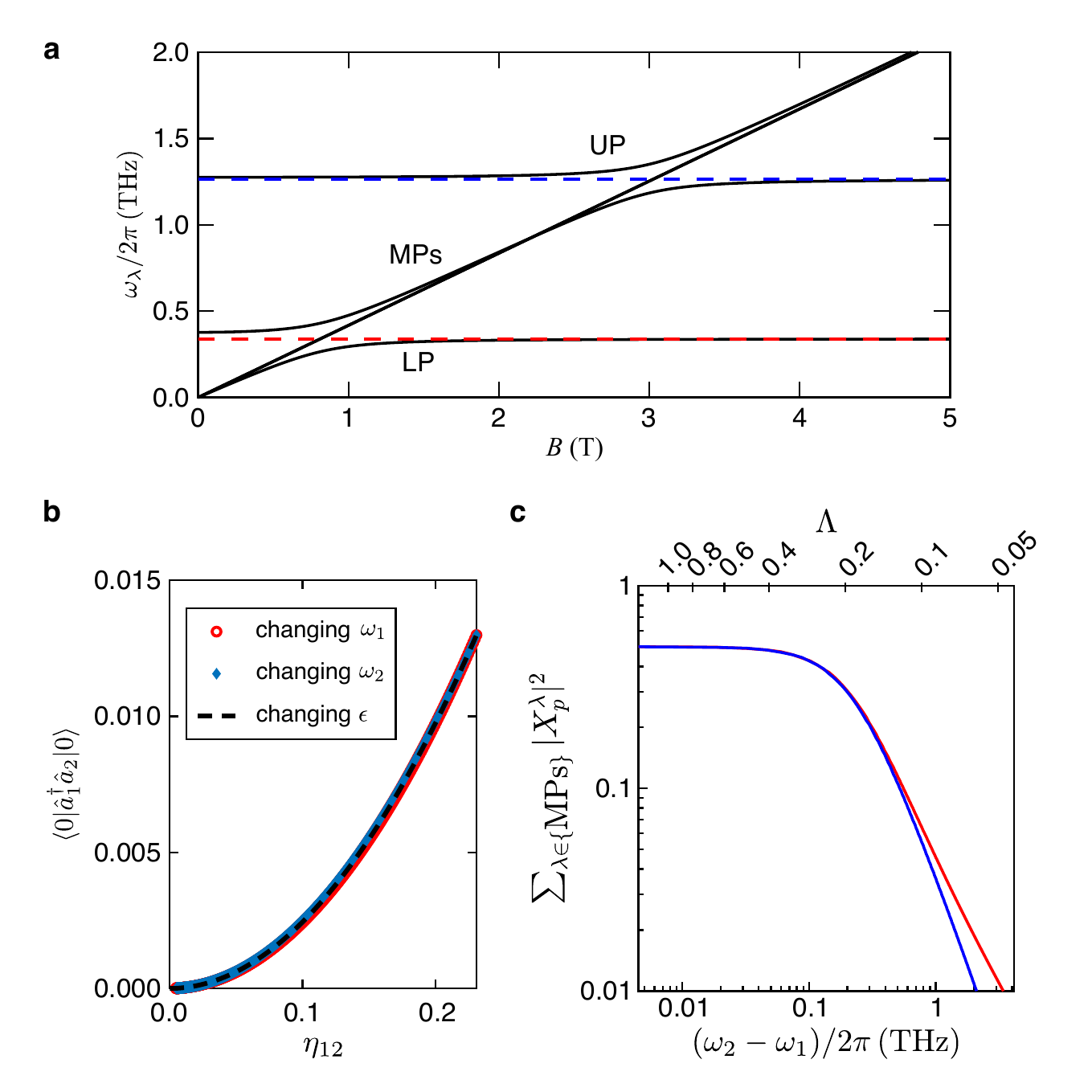}
\caption{\textbf{Polariton dispersion, ground-state correlations, and cavity photon weights of a toy model with two cavity modes and continuously tunable spatial overlap.} \textbf{a},~Polariton dispersion (black lines) as a function of the magnetic field $B$ in the case of perfect overlap of the cavity modes ($\epsilon=1$), showing an S-shaped MP. The red and blue dashed lines correspond to $\omega_{1}$ and $\omega_{2}$, respectively. \textbf{b},~Scaling of the intermode correlations with the off-diagonal coupling strength $\eta_{12}$. The red circles and blue diamonds are obtained by tuning the frequency $\omega_{1} =2\pi \times [0.339,1.8]$\,THz (while keeping $\omega_{2}=2 \pi \times 0.384$\,THz and $\epsilon=1$) and $\omega_{2}=2\pi \times [0.384,1.8]$\,THz (while keeping $\omega_{1}=2 \pi \times 0.339$\,THz and $\epsilon=1$), respectively. The black dashed line is obtained by tuning $\epsilon$ between 0 and 1, with $\omega_{1}=2 \pi \times 0.339$\,THz and $\omega_{2}=2 \pi \times 0.384$\,THz. The magnetic field is set to $B=0.81$\,T, for which $\omega_{c}=\omega_{1}$. The overlap of all traces confirms that the intermode correlations at the ground state are governed by the FOM $\eta_{12}$. \textbf{c},~Weights of the MPs, $\sum_{\lambda \in\{\textrm{MPs}\}} \vert X^{\lambda}_{p}\vert^{2}$, onto the different cavity modes $p$ as a function of the frequency difference $\omega_{2}-\omega_{1}$ with the SSC FOM $\Lambda$ on top. Here $\omega_{2}$ is increased while keeping $\omega_{1}=2\pi \times 0.339$\,THz and the overlap parameter $\epsilon=1$ fixed. For each $\omega_{2}$, the magnetic field is adjusted to the value corresponding to the inflexion point where a MP crosses the CR (see \textbf{a}).}
\label{fig:toymodel}
\end{figure}

The Hamiltonian \eqref{Hfin_A2_toy2} can be written in terms of the CR excitation operators $$\hat{b}_{\textbf{G}}=\int \! \frac{d\bm{\rho}}{a} \, \hat{b} (\bm{\rho}) e^{-i \textbf{G}\cdot \bm{\rho}},$$ decomposed over the set of reciprocal lattice vectors $G_{j}=m_{j}\times2\pi/a$ ($m_{j}=0,1,2,...$ and $j=x,y$) as 
\begin{align}
H&=\sum_{p} \hbar \omega_{p} \hat{a}^{\dagger}_{p} \hat{a}_{p} + \hbar\omega_\text{c} \sum_{\textbf{G}} \hat{b}^{\dagger}_{\textbf{G}} \hat{b}_{\textbf{G}} \nonumber \\
&+ \sum_{p,\textbf{G}} \hbar g_{p} (\textbf{G}) \left[\hat{b}_{-\textbf{G}} + \hat{b}^{\dagger}_{\textbf{G}} \right] \left(\hat{a}_{p} + \hat{a}^{\dagger}_{p}\right) \nonumber \\
& + \sum_{p,p'} \sum_{\textbf{G}} \frac{\hbar g_{p}(\textbf{G})g_{p'}(\textbf{G})}{\omega_\text{c}} \left(\hat{a}_{p} + \hat{a}^{\dagger}_{p}\right) \left(\hat{a}^{\dagger}_{p'} + \hat{a}_{p'}\right),
\label{Hfin_A2_G}
\end{align}
with $g_{p} (\textbf{G})=\int \! \frac{d\bm{\rho}}{a^{2}} \, g_{p}(\bm{\rho}) e^{-i \textbf{G}\cdot \bm{\rho}}$. The only non-vanishing Fourier coefficients $g_{p} (\textbf{G})$ of the mode profile functions \eqref{mode_prog} correspond to $m_{x}=\pm 1$ and $m_{y}=\pm 1$. The Hamiltonian \eqref{Hfin_A2_G} can be put in the diagonal form $H=\sum_{\lambda} \omega_{\lambda}\hat{p}^{\dagger}_{\lambda}\hat{p}_{\lambda}$, with the polariton modes
\begin{align}
\hat{p}_{\lambda} &= \sum_{p} X^{\lambda}_{p} \hat{a}_{p} + \sum_{\textbf{G}} W^{\lambda}_{\textbf{G}} \hat{b}_{\textbf{G}} + \sum_{p} \widetilde{X}^{\lambda}_{p} \hat{a}^{\dagger}_{p} + \sum_{\textbf{G}} \widetilde{W}^{\lambda}_{\textbf{G}} \hat{b}^{\dagger}_{-\textbf{G}}.
\label{bogol_tr}
\end{align}
Here $\lambda$ takes 6 values, including a LP (lowest eigenvalue), an UP (largest eigenvalue), and 4 intermediate eigenvalues (MPs) lying in between the LP and the UP (see Fig.~\ref{fig:toymodel}a).

By inverting the transformation \eqref{bogol_tr}, one finds that the correlations between the cavity modes in the state $\ket{\{n_{\lambda}\}}$ (containing $n_{\lambda}$ polaritons in the mode $\lambda$) read
\begin{align}
\bra{\{n_{\lambda}\}}\hat{a}^{\dagger}_{p} \hat{a}_{p'} \ket{\{n_{\lambda}\}}=\sum_{\lambda} \left(\widetilde{X}^{\lambda}_{p} \right)^{*} \widetilde{X}^{\lambda}_{p'} \left( n_{\lambda} +1 \right) + \sum_{\lambda} X^{\lambda}_{p}  \left(X^{\lambda}_{p'} \right)^{*} n_{\lambda}.
\label{corr_eq_re}
\end{align}
In the polariton vacuum, $n_{\lambda}=0$ $\forall \lambda$, the only contribution is the first term in the right-hand side of Eq.~\eqref{corr_eq_re}, which depends on the anomalous coefficients $\widetilde{X}^{\lambda}_{p}$. As a hallmark of the USC regime, the latter are expected to be governed by the USC figure of merit. In particular, we find that the intermode ground-state correlations are well captured by the figure of merit $$\eta_{12} = \sqrt{\frac{\int (d\bm{\rho}/a^{2}) \, g_{1}(\bm{\rho})g_{2}(\bm{\rho})}{\omega_\text{c}(\omega_{1}+\omega_{2})/2}}$$ introduced in the main text (Fig.~\ref{fig:toymodel}b). On the other hand, the correlations between the cavity modes in the excited states ($n_{\lambda} \neq 0$) also depend on the coefficients $X^{\lambda}_{p}$ (second term in the right-hand side of Eq.~\eqref{corr_eq_re}), which correspond to the weights of the polariton mode $\lambda$ onto the different cavity modes $p$. The weights of the MPs onto the modes $p=1,2$ are displayed in Fig.~\ref{fig:toymodel}c as a function of the frequency difference $\omega_{2}-\omega_{1}$, showing that the intermode correlations in the excited states are expected to be enhanced in the SSC.


\section{Contributions of the different terms in the full Hamiltonian to the ground-state correlations}
\begin{figure}[ht!]%
\centering
\includegraphics[width=0.95\textwidth]{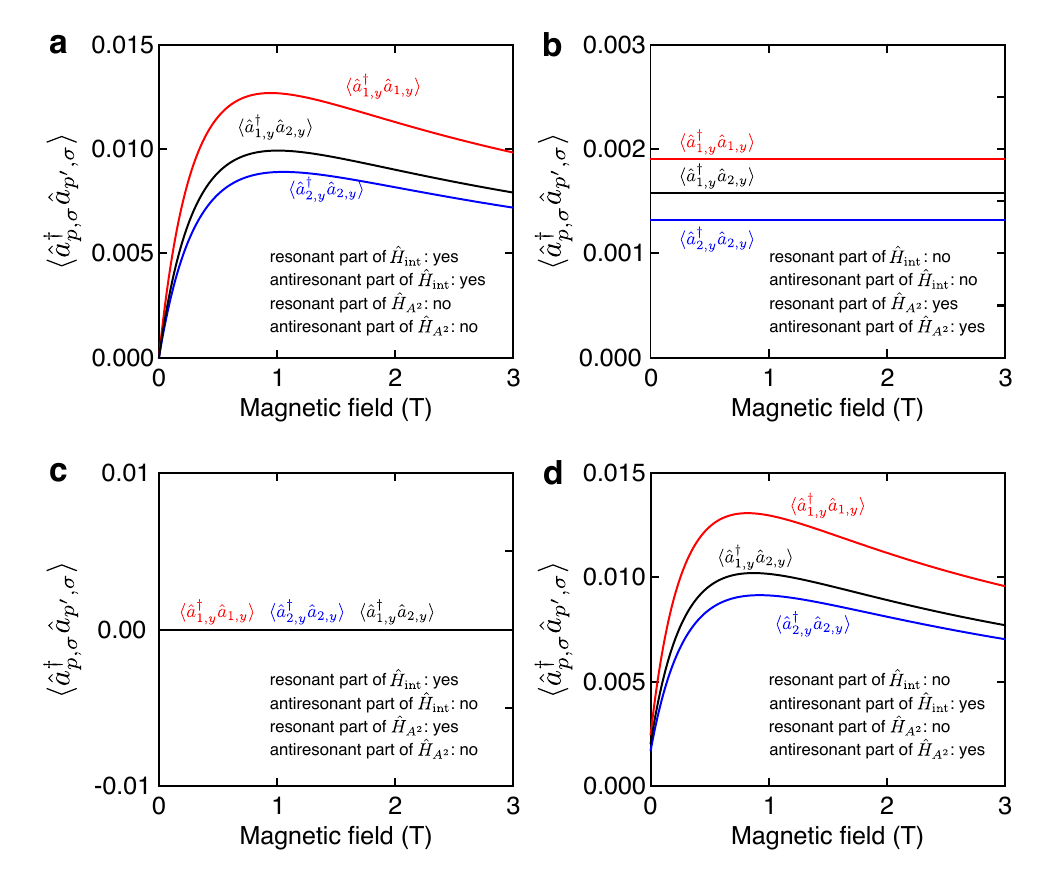}
\caption{\textbf{Contribution of the ground-state correlation from different terms in the full Hamiltonian.} \textbf{a--d},~Calculated ground-state correlations $\langle\hat{a}^{\dagger}_{p,j}\hat{a}_{p',j}\rangle$ for TE modes (\textbf{a}) with only the $\hat{H}_{\textrm{int}}$ term, (\textbf{b}) with only the $\hat{H}_{A^2}$ term, (\textbf{c}) with only the resonant part of the $\hat{H}_{\textrm{int}}$ and $\hat{H}_{A^2}$ terms, and (\textbf{d}) with only the anti-resonant part of the $\hat{H}_{\textrm{int}}$ and $\hat{H}_{A^2}$ terms.}
\label{fig:correlationdifferentterms}
\end{figure}

Figure~\ref{fig:correlationdifferentterms} shows the contribution of the ground-state correlations from different terms in the Hamiltonian. The $\hat{H}_{\textrm{int}}$ term provides the magnetic field dependence of the ground-state correlation (Fig.~\ref{fig:correlationdifferentterms}a). The $\hat{H}_{A^2}$ term provides the baseline of the ground-state correlation with the full Hamiltonian (Fig.~\ref{fig:correlationdifferentterms}b), supporting the claim that the $\eta_{pp',j}$ is a suitable FOM for ultrastrong photon--photon coupling. The ground-state correlations are zero if the antiresonant part of the $\hat{H}_{\textrm{int}}$ and $\hat{H}_{A^2}$ terms are neglected (Fig.~\ref{fig:correlationdifferentterms}c). The ground-state correlation will be overestimated if the resonant part of the $\hat{H}_{\textrm{int}}$ and $\hat{H}_{A^2}$ terms are neglected (Fig.~\ref{fig:correlationdifferentterms}d).

\section{Vacuum fluctuations and localization of the cavity modes}
\label{sec:vacuum fluc}

The vacuum fluctuations of the photonic modes shown in the main text were computed as follows: The expectation value of the electric energy density along $z$ in the vacuum state $\ket{0}$ is
$\langle I (z)\rangle = \int \! d\bm{\rho} \, \varepsilon_{0} \varepsilon (\bm{\rho},z) \sum_{j=x,y,z} \bra{0} \hat{E}^{2}_{j} (\bm{\rho},z) \ket{0}/2$, with the component $j$ of the electric field operator
\begin{align}
\hat{E}_{j} (\bm{\rho},z)=i\sum_{p,\sigma} \sqrt{\frac{\hbar \omega_{p}}{2\varepsilon_{0}a^{3}}} E_{p,\sigma,j} (\bm{\rho},z) \left(\hat{a}_{p,\sigma}- \hat{a}^{\dagger}_{p,\sigma} \right).
\end{align}
The contribution of the cavity mode $p$ with polarization $\sigma$ to that expectation value [i.e., $\langle I (z)\rangle=\sum_{p,\sigma} \langle I_{p,\sigma}(z) \rangle$] thus reads
\begin{align}
\langle I_{p,\sigma}(z) \rangle = \frac{\hbar \omega_{p}}{4a} \int \! \frac{d\bm{\rho}}{a^{2}} \, \varepsilon (\bm{\rho},z) \sum_{j=x,y,z} E^{2}_{p,\sigma,j} (\bm{\rho},z).
\end{align}
Note that the integral runs over a woodpile unit cell of surface $a^2=(333\,\upmu\textrm{m})^2$.

The variance of the in-plane electric field (at the vertical location of the 2DEG) in the vacuum state, $\langle E^{2}(\bm{\rho}) \rangle =\bra{0}\hat{E}^{2}_{x} (\bm{\rho},z_\text{2DEG}) \ket{0}+\bra{0}\hat{E}^{2}_{y} (\bm{\rho},z_\text{2DEG}) \ket{0}$, can be computed in a similar way, i.e., by writing $\langle E^{2}(\bm{\rho}) \rangle =\sum_{p,\sigma} \langle E^{2}_{p,\sigma}(\bm{\rho}) \rangle$, with
\begin{align}
\langle E^{2}_{p,\sigma}(\bm{\rho}) \rangle= \frac{\hbar \omega_{p}}{2 \varepsilon_{0} a^{3}} \left[E^{2}_{p,\sigma,x} (\bm{\rho},z_\text{2DEG}) + E^{2}_{p,\sigma,y} (\bm{\rho},z_\text{2DEG})\right].
\end{align}
The quantity plotted on Fig.~2c--f is the standard deviation $\sqrt{\langle E^{2}_{p,\sigma}(\bm{\rho}) \rangle}$.

\begin{figure}[ht!]%
\centering
\includegraphics[width=0.95\textwidth]{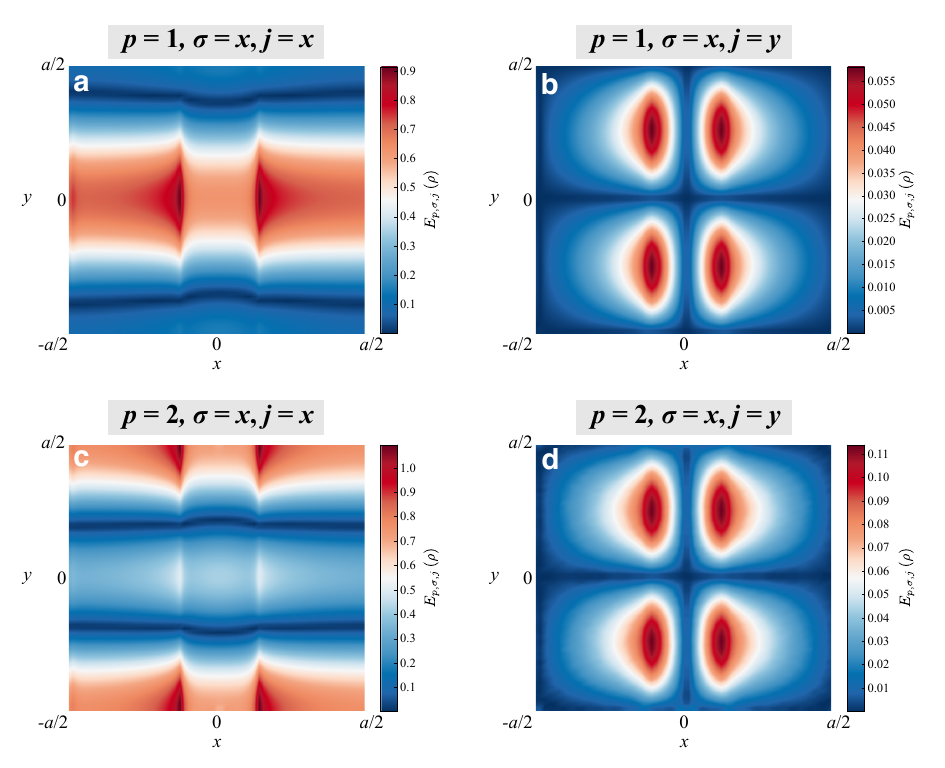}
\caption{\textbf{Real-space distribution of electric field in different components.} \textbf{a--d},~The $j=x$ and $j=y$ components of the electric field profile of cavity mode $p=1$ (\textbf{a--b}) and mode $p=2$ (\textbf{c--d}) for $\sigma=x$.}
\label{fig:ExEymodeprofiles}
\end{figure}

The $j=x$ and $j=y$ components of the variance of the in-plane electric field in the vacuum state are displayed in Fig.~\ref{fig:ExEymodeprofiles}. It is worthy to note that $\sqrt{\langle E^{2}_{p,\sigma}(\bm{\rho}) \rangle}$ is mainly polarized along the $\sigma$ direction. The mode profiles for $\sigma\neq j$ (Fig.~\ref{fig:ExEymodeprofiles}b,d) are much weaker than those for $\sigma=j$ (Fig.~\ref{fig:ExEymodeprofiles}a,c).

In order to assess the degree of localization of the different cavity modes in the $z$ direction, one can treat the electric energy density of the mode $p$ with polarization $\sigma$ as a probability distribution since
\begin{align}
\int \! \frac{d\bm{\rho}}{a^{2}} \int \! \frac{dz}{a} \, \varepsilon (\bm{\rho},z) \sum_{j=x,y,z} E^{2}_{p,\sigma,j} (\bm{\rho},z) =1,
\end{align}
and compute the standard deviation of the electric energy density along $z$ as
\begin{align}
\sigma_{p,\sigma}=\sqrt{\int \! \frac{d\bm{\rho}}{a^{2}} \! \int \! \frac{dz}{a} \, (z-\bar{z}_{p,\sigma})^{2} \, \varepsilon (\bm{\rho},z) \sum_{j=x,y,z} E^{2}_{p,\sigma,j} (\bm{\rho},z)},
\end{align}
with the mean value
\begin{align}
\bar{z}_{p,\sigma}=\sqrt{\int \! \frac{d\bm{\rho}}{a^{2}} \! \int \! \frac{dz}{a} \, z \, \varepsilon (\bm{\rho},z) \sum_{j=x,y,z} E^{2}_{p,\sigma,j} (\bm{\rho},z)}.
\end{align}
Since the photonic modes are localized in the vicinity of the defect layer along the $z$ direction and exhibit discrete translational symmetry in the plane, we compute the spatial profile $E_{p,\sigma,j} (\bm{\rho},z)$ in a woodpile unit cell. The integral along $z$ runs over the whole computational cell, which includes air layers on each side of the woodpile structure along the $z$ direction. The standard deviations of the 3D-PCC are $\sigma_{1,x}\approx \sigma_{2,x} \approx \sigma_{1,y} \approx \sigma_{2,y}\approx 0.5a$ (about two Si logs thick), and $\sigma_{3,x} \approx \sigma_{4,x} \approx \sigma_{3,y}\approx \sigma_{4,y}\approx 0.18a$ (less than one Si log thick). For all cavity modes, more than $95 \%$ of the electric energy is located in the 3D-PCC. As expected, the localization of the cavity modes along the $z$ direction increases as the latter are located deeper into the photonic band gap.

\section{Numerical simulations}
\label{sec:COMSOL simulations}
The mode profiles of the cavity modes induced by a 60\,$\upmu$m-thick bare GaAs layer were calculated by the MEEP software~\cite{OskooiEtAl2010CPC}. The electric field was assumed to be constant over the thickness $d\simeq 2$\,$\upmu$m of the QW heterostructure. The transmittance spectrum of the bare cavity in the main text was simulated by using the COMSOL Multiphysics software. We used the permittivity $\varepsilon=11.6964$ for the silicon layers, and $\varepsilon=12.96$ for the GaAs layer. For the simulations with the 2DEG layer in the main text, because the woodpile structure exhibits mirror symmetries, only 1/4 of the unit cell of the woodpile structure was considered in the geometry to reduce the simulation time. A transition boundary condition with an effective thickness, $d=2$\,$\upmu$m, was used to emulate the MQW structure. A gyrotropic permittivity tensor was used to describe the complex permittivity of the 2DEG layer at different magnetic fields:
\begin{equation}
    \tilde{\varepsilon} = \begin{pmatrix} \varepsilon_{xx}(\omega) & \varepsilon_{xy}(\omega) & 0 \\ -\varepsilon_{xy}(\omega) & \varepsilon_{xx}(\omega) & 0 \\ 0 & 0 & \varepsilon_{zz} \end{pmatrix},
\end{equation}
with
\begin{align}
\varepsilon_{xx}(\omega) &= \varepsilon_{\text{bg}} - \frac{\omega_\text{pl}^2(\omega-i\gamma)}{\omega d[(\omega-i\gamma)^2-\omega_\text{c}^2]},\\
\varepsilon_{xy}(\omega) &= \frac{-i\omega_\text{pl}^2\omega_\text{c}}{\omega d[(\omega-i\gamma)^2-\omega_\text{c}^2]},\\
\varepsilon_{zz} &= \varepsilon_{\text{bg}},
\label{eq22}
\end{align}
where $\varepsilon_{\text{bg}}=12.96$ is the background permittivity, $\omega_\text{pl}=\sqrt{n_e e^2/(\varepsilon_0 m_{\textrm{eff}})}$ is the plasma frequency and $\omega_\text{c}= e B/m_{\textrm{eff}}$ is the cyclotron frequency.

\section{Thickness of the defect layer in numerical calculations}
\begin{figure}[ht!]%
\centering
\includegraphics[width=0.5\textwidth]{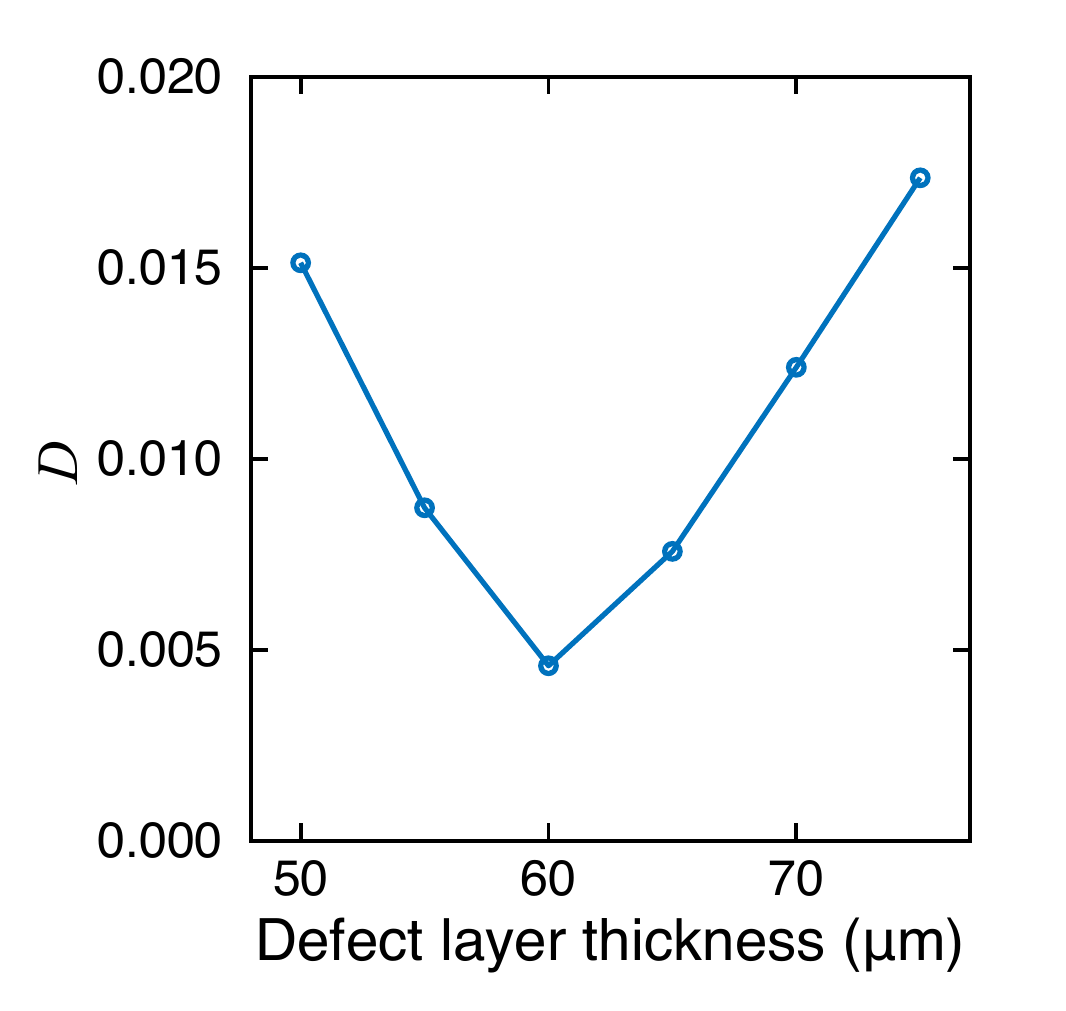}
\caption{\textbf{Optimization of the thickness of the defect layer.} Deviation, $D$, as a function of the defect layer thickness. $D$ reaches its minimum when the defect layer is 60-$\upmu$m-thick.}
\label{fig:defectlayer thickness}
\end{figure}

Due to imperfections in mechanical polishing, the thickness of the GaAs substrate of the 2DEG layer in the experiment was not perfectly homogeneous. To determine the suitable thickness of the substrate for numerical simulations, we ran simulations in the presence of the 2DEG layer at $B = 0,\,1.5,\,2.5,\, \text{and}\, 7$\,T with different thicknesses of the GaAs substrate. We calculated the deviations of the polariton frequencies for TE and TM modes between the experiment and simulation, $\Delta f_{P,B}= f_{\text{exp}}-f_{\text{sim}}$, where $P=\{\text{UP},\text{LP1},\text{LP2}\}$ at each $B$. Finally, we calculated the root mean square of the total deviation,
\begin{equation}
    D = \sqrt{\frac{\sum_{\text{TE},\text{TM}}\sum_{B}\sum_{P}\Delta f_{P,B}^2}{2N}},
\end{equation}
where $N=13$ is the number of data points.
The optimized thickness of the defect layer that gives the minimum deviation is 60 $\upmu$m (including the 2-$\upmu$m-thick MQW layer), as shown in Fig.~\ref{fig:defectlayer thickness}. This is consistent with the profilometer measurement.

\section{Simulations with different ranges of time delays}
\begin{figure}[ht!]%
\centering
\includegraphics[width=0.8\textwidth]{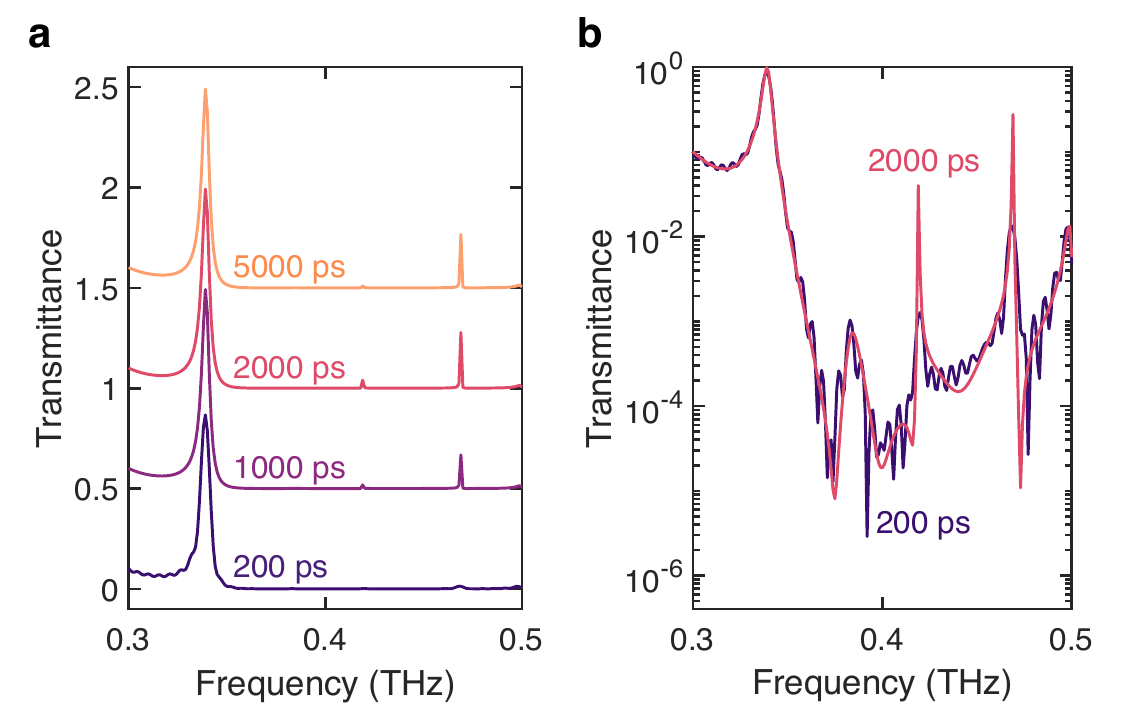}
\caption{\textbf{Simulations with different lengths of time window.} \textbf{a},~Transmittance spectra of a bare 3D-PCC with a series of time windows obtained from simulations. The peaks corresponding to the third and fourth cavity modes become pronounced when the time window is longer than 1000\,ps. \textbf{b},~Transmittance spectra in (\textbf{a}) that are plotted in logarithmic scale.}
\label{fig:Lumerical timewindows}
\end{figure}

We investigated the transmittance spectra of the system with different ranges of time delays in FDTD simulations (Lumerical). As shown in Fig.~\ref{fig:Lumerical timewindows}, in order to resolve modes 3 and 4 with high $Q$-factors in the spectrum, a long time-domain trace is needed. The amplitudes of modes 3 and 4 are weak for a time-domain trace up to 200\,ps.

\section{Extraction of peak frequencies}
\label{sec:extract freq}
\begin{figure}[ht!]%
\centering
\includegraphics[width=0.8\textwidth]{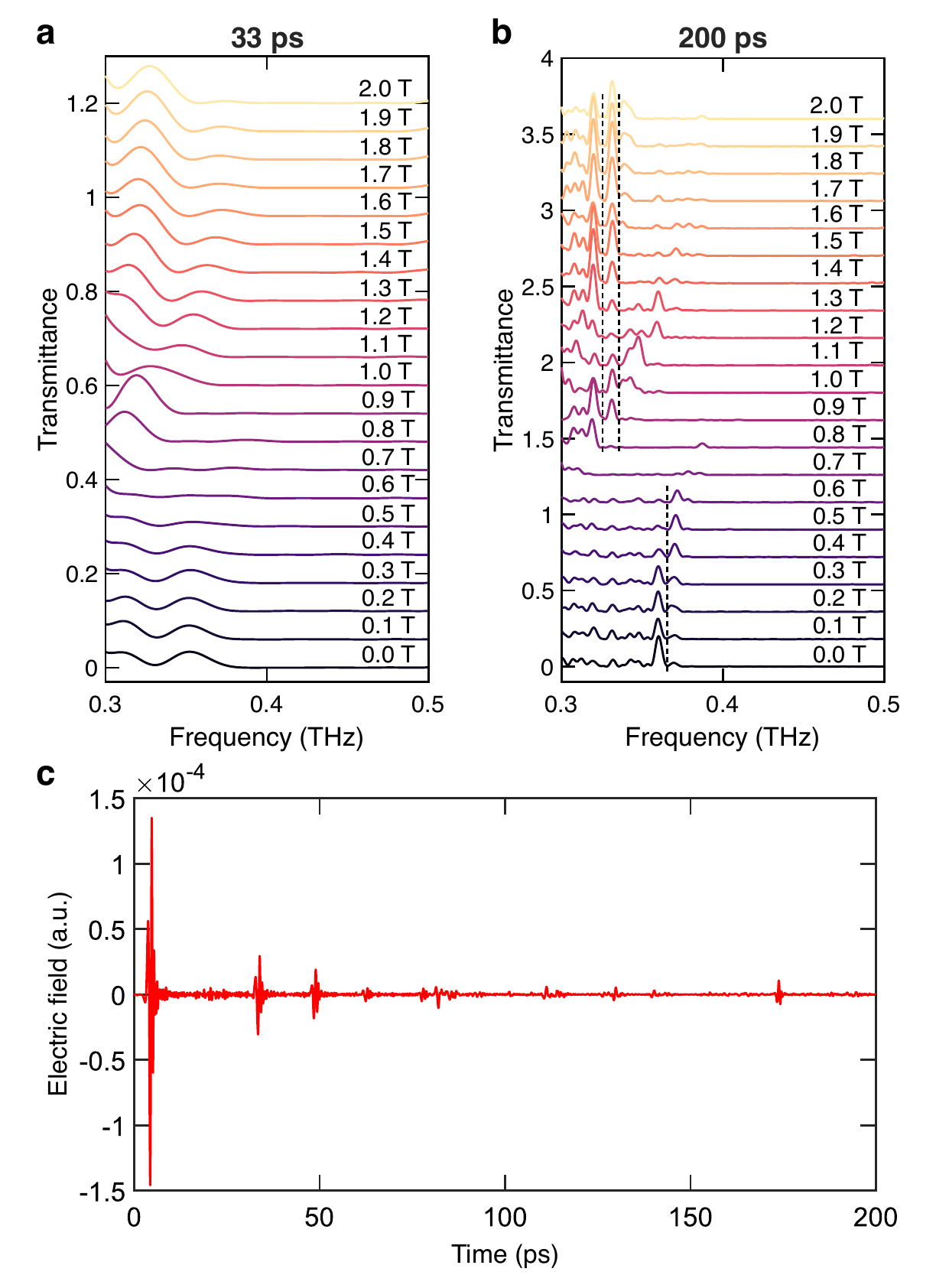}
\caption{\textbf{Experimental data with different lengths of the time window.} \textbf{a--b},~Transmittance spectra with a time window up to 33\,ps (\textbf{a}) and 200\,ps (\textbf{b}) for TM modes. The black dashed lines in (\textbf{b}) mark the dips induced by Fabry-P\'{e}rot modulation, which are insensitive to a magnetic field. \textbf{c},~Temporal waveform of THz radiation without passing through a sample. Echoes of THz pulses come from multiple optical components in the system, for instance, cryostat windows and crystals.}
\label{fig:time windows}
\end{figure}

\begin{figure}[ht!]%
\centering
\includegraphics[width=0.9\textwidth]{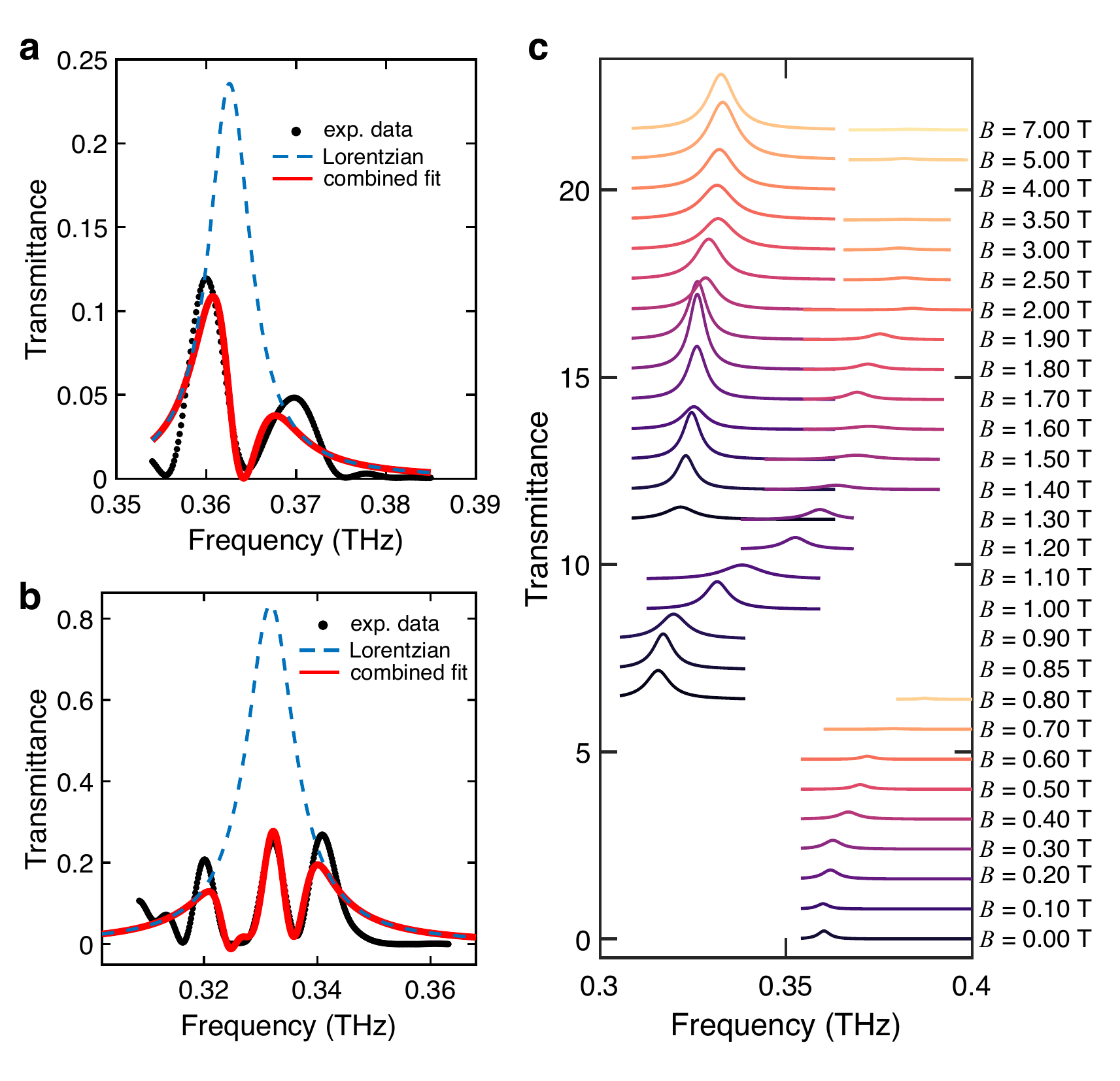}
\caption{\textbf{Extraction of peak frequencies from transmittance spectra with a long time-domain waveform.} \textbf{a--b},~Fittings to experimental transmittance spectra with a combination of a Lorentzian peak and Gaussian dips for $B=0.3$\,T (\textbf{a}) and $B=7$\,T (\textbf{b}) for TM modes. \textbf{c},~The Lorentzian peak extracted from the fittings as a function of magnetic field.}
\label{fig:Lorentzian peak}
\end{figure}

The linewidth of the polariton branches in the color plot in the main text is limited by the frequency resolution of the measurements as a short time-domain range (33\,ps) was considered. Scanning a long range of time delays is required for THz-TDS measurements to resolve the actual linewidth of the coupled modes in the high-$Q$ 3D-PCC. For instance, Fig.~\ref{fig:time windows}a,b display experimental transmission spectra with different ranges of time delays (33\,ps and 200\,ps) for TM modes. As the length of the time-domain trace increases, the linewidths of the polaritons in the experimental data after applying Fourier transform become narrower. However, echoes of THz pulses from the optical components are unavoidable in the collected data at longer delays. The echoes of THz radiation were induced by multiple components in the system, e.g., cryostat windows and crystals, and thus, the time separation between the pulses is not constant, as shown in Fig.~\ref{fig:time windows}c. This leads to multiple dips, which are not equally spaced, in the transmittance spectra of the sample (i.e., Fabry--P\'{e}rot effect) when a long time-domain trace is used. It causes the artificial splitting of the peaks when the peaks in the spectra are superimposed on the dips. Note that the dips caused by the Fabry--P\'{e}rot effect can be identified because the dip frequencies did not change with $B$ (black dashed lines in Fig.~\ref{fig:time windows}b). To extract the peak position from the spectra with Fabry-P\'{e}rot modulation, we fitted the peaks with a combination of a Lorentzian peak and a few Gaussian dips. For example, Fig.~\ref{fig:Lorentzian peak}a,b shows the fittings for TM modes at $B=0.3$\,T and $B=7$\,T, respectively. The Lorentzian peaks extracted from the fittings at different magnetic fields for the TM modes are shown in Fig.~\ref{fig:Lorentzian peak}c. The peak frequencies of the Lorentzian peak are shown as the white dots in the main text. The white dots extracted from long time-domain traces show agreement with the color plot extracted from short time-domain traces.

\clearpage

\bibliography{USC}